\documentclass[a4paper,11pt]{article}
\usepackage{epsfig}
 \usepackage{natbib}
\usepackage{algorithm}
\usepackage{algorithmic}
\usepackage[latin1]{inputenc}
 \usepackage[T1]{fontenc}
 \usepackage[normalem]{ulem}
 \usepackage[english]{babel}
 \usepackage{verbatim}
 \usepackage{graphicx}
\usepackage{algorithmic} 
\usepackage{algorithm} 
\usepackage{amsthm, amsmath, amssymb, amsfonts}
\usepackage[modulo,mathlines]{lineno}
\usepackage{array} 

\usepackage{latexsym,amsfonts,amsmath,amssymb}

\begin{document}
\title{Capturing the time-varying drivers of an epidemic using stochastic dynamical systems}

\author{JOSEPH {DUREAU}$^\ast$, KONSTANTINOS {KALOGEROPOULOS}\\[4.1pt]
{\it Department of Statistics , London School of Economics}\\
{\it and Political Science, UK{}}\\[2pt]
{j.dureau@lse.ac.uk}\\
MARC  {BAGUELIN}\\[4.1pt]
{\it Immunisation, Hepatitis and Blood Safety Department,} \\
{\it Health Protection Agency, }\\
{\it Centre for the Mathematical Modeling of Infectious} \\ 
{\it  Diseases, London School of Hygiene and Tropical Medicine}}

\maketitle

\markboth{J. Dureau, K. Kalogeropoulos and M. Baguelin}{Capturing the time-varying drivers of an epidemic}

\maketitle

\footnotetext{To whom correspondence should be addressed.}

\begin{abstract}
{
Epidemics are often modelled using non-linear dynamical systems observed through partial and noisy data. In this paper, we consider stochastic extensions in order to capture unknown influences (changing behaviors, public interventions, seasonal effects etc). These models assign diffusion processes to the time-varying parameters, and our inferential procedure is based on a suitably adjusted adaptive particle MCMC algorithm. The performance of the proposed computational methods is validated on simulated data and the adopted model is applied to the 2009 H1N1 pandemic in England. In addition to estimating the effective contact rate trajectories, the methodology is applied in real time to provide evidence in related public health decisions. Diffusion driven SEIR-type models with age structure are also introduced.
}
{population epidemic model; time-varying parameters; Bayesian inference;  Particle MCMC}
\end{abstract}

\section{\label{sec1}Introduction}

Epidemic models are often used to simulate disease transmission dynamics, detect emerging outbreaks \citep{Unkel2012}, and assess public health interventions \citep{Boily2007}. In order to capture the dynamics of epidemics, the main focus is generally made on their intrinsically dynamic elements such as the depletion of susceptibles or the population immunity evolution. Nevertheless, there are time-varying extrinsic factors that are crucial to the epidemic course. These may include social cycles (holidays), public interventions and climatic variations. This has been illustrated for diseases such as cholera, malaria \citep{Cazelles2005,Ionides2006} or influenza \citep{Shaman2009}. These studies were conducted either by relating climatic and incidence time-series \citep{Cazelles2005}, which does not disentangle the effect of intrinsic and extrinsic factors, or by experimentally assessing the virus resistance in different climatic conditions \citep{Shaman2009} requiring an extrapolation to the population scale. Overall, the time-varying nature of epidemics poses a challenging statistical problem stressing the need for suitable computational tools \citep{Ferguson2007}.

This paper considers a flexible modelling framework that encompasses time-varying aspects of the epidemic via stochastic differential equations. We aim at providing robust inferential procedures, incorporating the uncertainty associated with key parameters and accounting for data and model limitations. In order to provide an accurate and feasible computational toolbox, we provide Markov Chain Monte Carlo (MCMC) algorithms utilising recent developments such as particle MCMC (PMCMC) algorithms \citep{Andrieu2010} and adaptive techniques \citep{Roberts2009}. Modelling aspects are presented in Section 2, while the computational framework is presented in Section 3. In Section 4 we evaluate the performance of the proposed adaptive PMCMC schemes on simulated data. In Section 5 we present various applications of the methodology to the 2009 A/H1N1 pandemic, and conclude, in Section 6, with some relevant discussion. Further simulations can be found in the Supplementary Materials.

\section{Modelling framework}

\subsection{Epidemic models with time-varying coefficients}
\label{ssec:EpiModels}

We adopt a SEIR model as a guide in this paper, although the methodology can be applied to other dynamical systems. The model is set in \eqref{eq:SEIR}; S accounts for susceptible, E for infected but not infective, I for infective, and R for removed individuals. New infections occur at a rate $\beta S_{t}\frac{I_{t}}{N}$, implying that the susceptible individuals make effective contacts at rate $\beta$ (the effective contact rate), and only a fraction $\frac{I_{t}}{N}$ of these contacts are made with infective individuals. The average period spent in compartments $E$ and $I$ is given by $k^{-1}$ and $\gamma^{-1}$ respectively.
\begin{equation}
\label{eq:SEIR}
\frac{dS_{t}}{dt} =-\beta S_{t}\frac{I_{t}}{N},\;\;\;\; \frac{dE_{t}}{dt} =\beta S_{t}\frac{I_{t}}{N}-kE_{t},\;\;\;\;\frac{dI_{t}}{dt} =kE_{t}-\gamma I_{t},\;\;\;\; \frac{dR_{t}}{dt} =\gamma I_{t}
\end{equation}
The basic reproduction number, $R_{0}$, represents the number of secondary infections from a primary infected individual in a fully susceptible population. A related quantity is the effective reproduction number, $R_{t}$, refers to the number of secondary cases from an infected individual at time $t$. $R_{t}$ is a context-dependent quantity of high interest to policy makers as it indicates the possibility for the epidemic to grow ($R_{t}>1$) or to decrease ($R_{t}<1$) \citep{Anderson1992}.

Epidemic models can be quite detailed (including individual characteristics, geographic information etc.) or basic, such as the SEIR model, that geographically aggregates the cases and assumes deterministic transmission processes, occurring at a given frequency each time infected and susceptible meet. The latter are easier to estimate and interpret, but are based on strong assumptions that could lead to poor inference. In this paper we adopt stochastic extensions of the deterministic SEIR models. The additional dynamic error is likely to contain structural mis-specifications and can subsequently be explored and potentially revised. We focus on large-scale epidemics, for which random effects in transmission processes can be considered to be well-approximated deterministically \citep{Kurtz1981}. We adopt the paradigm that attributes the model limitations mainly to the time varying nature of the effective contact rate, henceforth denoted as $\beta_t$, rather than to the variability in individual characteristics or in transmission processes.

An early approach to estimate $R_{t}$ can be found in \cite{Fine1982}. It can be implemented through discrete generation models or by reconstructing the chain of transmission \citep{Cauchemez2006,Griffin2011}. However, as $R_{t}$ estimates contain both the effects of evolving transmissibility and immunity, quantitative conclusions can hardly be generalised to situations where the immunological situation is different. We therefore concentrate on estimating $\beta_{t}$ rather than $R_{t}$. A number of approaches use a finite-dimension function space for the trajectory of $\beta_{t}$. Low-dimensional examples can be found in \cite{Cauchemez2008}, in which $\beta_{t}$ is modeled as a piece-wise linear function. In some higher-complexity models, as in \cite{Cauchemez2008a} and \cite{Ionides2006}, $\beta_{t}$ is estimated freely with a few-weeks resolutions. Loosely speaking, as the number of parameters for the trajectory of $\beta_t$ increases, model-induced biases fade out at the expense of the variance. A compromise is required to improve robustness and is often controlled through a regularising parameter. For example, in \cite{He2011}, $\beta_{t}$ is estimated using cubic splines, calibrated via AIC.

\subsection{Diffusion driven epidemic models}
\label{ssec:DiffEpidModels}
We consider models where diffusion processes are used for some of the coefficients in (\ref{eq:SEIR}). Although alternative formulations are possible, as discussed in Section \ref{ssec:EpiModels}, we focus on $\beta_t$ to get
\begin{equation}
\begin{cases}
\frac{dS_{t}}{dt} &=-\beta_{t}S_{t}\frac{I_{t}}{N},\;\;\;\; \frac{dE_{t}}{dt} =\beta_{t}S_{t}\frac{I_{t}}{N}-kE_{t},\;\;\;\;\frac{dI_{t}}{dt} =kE_{t}-\gamma I_{t},\;\;\;\; \frac{dR_{t}}{dt} =\gamma I_{t},\\
dx_{t} & =\mu_x(x_t,\theta_x)dt +\sigma_{x}(x_t,\theta_x)dB_{t}, \;\;\;\;x_t=h(\beta_t),
\end{cases}\label{eq:SEIR-1}
\end{equation}
where $\mu_x(\cdot)$ denotes the drift, $\sigma_{x}(\cdot)$ the volatility and $h(\cdot)$ is a positive-valued function. The assigned diffusion may capture features such as behaviour changes, preventive measures, seasonal effects, holidays etc. When prior knowledge on $\beta_t$ is available, it can be reflected in $\mu_x(\cdot)$ and $\sigma_{x}(\cdot)$; e.g. if the contact rate is expected to converge, an Ornstein Uhlenbeck process can be chosen. Other options may include a sigmoid or a sinusoidal form; see for example \citep{Rasmussen2011}. In absence of prior information or when the researcher wants to impose little restrictions, a Brownian motion can be used, with $\mu_x(\cdot)\equiv0$ and $\sigma_{x}(\cdot)\equiv\sigma$ (i.e. $\theta_x=\sigma$). This model, with $h(\cdot)\equiv\log(\cdot)$, is henceforth denoted as BM. The obtained output can be either reported or used as an exploratory tool to construct a more structured model; see Section \ref{ssec:multiplegroups} for an application. The choice of BM implies a continuous, yet non-differentiable, path satisfying the Markov property. In cases where $\beta_t$ is believed to evolve as a smooth function in time, higher order Brownian motions could be used. Loosely speaking, these may be regarded as equivalent to non-parametric approaches such as cubic splines \citep{Wahba1990}, with the model in (\ref{eq:SEIR-1}) imposing a prior on $\beta_t$ and $\sigma$ being a regularising factor. The rate $\beta_t$ can be perceived as a product of a smooth and a rough component; the former being a population average of the intrinsic transmission procedure and latter containing extrinsic factors such as the amount of contact among individuals. It is therefore important to build a framework that contains both smooth and rough models.

The above model can be estimated with an Extended Kalman Filter (EKF), as in \cite{Cazelles1997}. EKF allows for fast computations, but is based on Taylor and Gaussian approximations whose error could be non-negligible; see Supplementary Materials for a relevant simulation experiment. Nevertheless, the EKF can still be used as a tool to construct efficient proposal distributions for MCMC schemes. It can also be used to optimize sequential Monte Carlo (SMC) algorithms, but either at a strong computational cost \citep{Sarkka2008b} or crude time discretisations \citep{Dukic2009}. Next, we develop a general framework for
efficient MCMC schemes that allow for good approximations.


\section{Data augmentation via MCMC for diffusion driven epidemic models}

This section presents a general inferential framework for diffusion-driven epidemic models. We adopt the Bayesian paradigm to incorporate parameter uncertainty and prior information in the estimates of $\beta_t$ trajectories. The problem can also be cast as estimating partially observed hypoelliptic diffusions, thus presenting various difficulties \citep{Pokern2009}. We begin by setting the model and justifying the need for data augmentation. Existing MCMC algorithms are considered but they can lead to extremely inefficient MCMC chains. We address the issue by taking advantage of the specific model structure to construct adaptive PMCMC schemes.

\subsection{Model and data augmentation setup}

For ease of exposition we focus on models satisfying (\ref{eq:SEIR-1}), but the framework covers models with different ODE systems or more time-varying coefficients, as in Section \ref{ssec:multiplegroups}. Being in continuous time, $t$ can take any value between $t_{0}$ and $t_{n}$. We denote the path of the ODE states vector $V_t=\{S_{t},E_{t},I_{t},R_{t}\}$ between observation times $t_{i}$ and $t_{j}$ by $V_{i:j}$. The data, $y_{1:n}=\{y_{t_{1}},..,y_{t_{n}}\}$, usually provide information for $I_t$ at specific times (prevalence data) or for integrals of $V_t$ (incidence data). In either case, we assume that they are obtained with error as the collection procedure is typically associated with additional uncertainty. The noise distribution is denoted with $\mathbb{P}_{y}$ with density $f(y_{1:n}|V_{0:n},\theta_y)$. Note that, in the model of (\ref{eq:SEIR-1}), $V_t$ can be written as a deterministic function, $g(\cdot)$, of $x_t$ and the parameters $\theta_v=(k,\gamma,V_0)$. This function is the solution of the ODE and can be written as an intractable time integral involving $x_{t}$. Hence, the model becomes
\begin{equation}
\begin{cases}
dx_{t} & =\mu_{x}(x_{t},\theta_x)dt+\sigma_{x}(x_t,\theta_x)dB_{t}\\
y_{1:n}| V_{0:n},\theta_y &\sim \mathbb{P}_{y}(y_{1:n}|V_{0:n},\theta_y),\;\;\;\;V_{0:n} =g(x_{0:n},\theta_v)
\end{cases}\label{eq:PartObsDiffusion}
\end{equation}
Denote with $\mathbb{P}_{x}$ the distribution of the diffusion $x_t$ defined from the SDE above. We require the existence of a unique weak solution which translates into some mild assumptions on $\mu_{x}(.)$ and $\sigma_x(.)$; e.g. locally Lipschitz with a linear growth bound, see for example \cite{Oksendal2003}. The distribution of $\mathbb{P}_{x}$ may also be viewed as a prior on $x_t$, or else $\beta_t$. The model can now be defined from $\mathbb{P}_{y}$, $\mathbb{P}_{x}$, and the assigned priors on $\theta=\{\theta_y,\theta_v,\theta_x\}$, denoted by $\pi(\theta)$
\begin{equation}
\label{eq:posterior}
\pi(x_{0:n},\theta|y_{1:n})\propto f(y_{1:n}|V_{0:n},\theta_y) \times d\mathbb{P}_{x} \times \pi(\theta)
\end{equation}
Given direct observations on $x_t$, it would have been possible to draw approximation-free inference on $d\mathbb{P}_{x}$ using the approach of \cite{Beskos2006}. However, this is not possible in our
case given the non-linear functionals in $g(\cdot)$ that render (3.4) intractable.We proceed by discretizing the path of $x_t$, and therefore of $\beta_t$ and $V_t$. More specifically, we introduce $m$ points between each pair of successive observation times $t_i$ and $t_{i+1}$ ($i=0,1,\dots,n-1$). When referring to the discrete representation of a path, the superscript $dis$ will be used; for example for a step $\delta=\frac{1}{m+1}$, the discrete skeleton of $x_{t}$ will be denoted by $x_{0:n}^{dis}=\{x_0,x_\delta,x_{2\delta},\dots,x_{t_n}\}$. The presence of $x_{0:n}^{dis}$ allows for approximations of (\ref{eq:posterior}) through the Euler-Maruyama scheme to evaluate $d\mathbb{P}_{x}$
\begin{equation}
\label{eq:apprPost}
\begin{cases}
p(x_{\delta:n}^{dis}|x_0,\theta_x)\;=\;\prod_{i:\;t_0<i\delta\leq t_n}p(x_{i\delta}|x_{(i-1)\delta},\theta_x),\\
x_{i\delta}|x_{(i-1)\delta}\; \sim \;\mathcal{N}\left\{x_{(i-1)\delta}+\delta\mu_{x}(x_{(i-1)\delta},\theta_x), \; \delta\sigma_{x}(x_{(i-1)\delta},\theta_x)^2\right\}.
\end{cases}
\end{equation}
Moreover, given $x_{0:n}^{dis}$,the ODE can be solved numerically to obtain $V_{0:n}^{dis}$ and evaluate $f(\cdot)$. The approximation error can be made arbitrarily small by increasing the user-specified parameter $m$.

\subsection{Data augmentation via Gibbs schemes}
\label{ssec:gibbs}

Model (\ref{eq:PartObsDiffusion}) can be put in the context of \cite{Chib2006}, \cite{Golightly2008} or \cite{Kalogeropoulos2007}. In these approaches, a Gibbs scheme can be used to sample from the joint posterior in (\ref{eq:posterior}) of $x_{0:n}^{dis}$ and $\theta$. The data augmentation algorithm alternates between drawing $x_{0:n}^{dis}$ given $\theta$, and updating $\theta$ conditional on the augmented path $x_{0:n}^{dis}$. The MCMC protocol ensures that the chain provides samples from the marginal posteriors of $x_{0:n}^{dis}$ and $\theta$. Nevertheless, the properties  of the algorithm may become unacceptably poor. There are two essential issues associated with such schemes. The first concerns the non-trivial step of sampling on the diffusion pathspace of $x_t$. The second problem is caused by the high posterior correlations between  $x_{0:n}^{dis}$ and $\theta$, leading to reducible chains as $m$ increases \citep{Roberts2001}.

The majority of the literature on data augmentation schemes for diffusions handles the conditional updates of $x_{0:n}^{dis}$ with an independence sampler. As it is difficult to find good proposal distributions for the entire $x_{0:n}^{dis}$, the path is usually split into blocks. Overlapping blocking strategies are essential to ensure that all points are updated and continuity of the path is retained. An alternative way to update $x_{0:n}^{dis}$ is to use the particle filter via the Particle Gibbs algorithm of \cite{Andrieu2010}. But unless the issue of high posterior correlation between $x_{0:n}^{dis}$ and $\theta$ is resolved, none of these schemes will improve the overall MCMC performance. The problem is caused by the quadratic variation process of $x_t$ that identifies $\theta_x$. For $\sigma_x(x_s,\theta_x)\equiv \sigma$ we get
\begin{equation}
\label{eq:quadraticvariation}
\lim_{\delta\rightarrow 0}\sum_{i:\; t_0 <i\delta \leq t_n} (x_{i\delta}-x_{(i-1)\delta})^2=\int_{t_0}^{t_n}\sigma^2ds=\sigma^2 (t_n-t_0)
\end{equation}
Thus, the conditional posterior of $\sigma$ converges to a point mass as $\delta$ tends to $0$. In practice this translates into an increasingly slow MCMC algorithm with a convergence rate of $O(m)$ \citep{Roberts2001}. Schemes
with a fixed $m$ \citep{Cori2009} could work in some occasions but the approximation error could be substantial. In some cases, the problem can be tackled with suitable reparametrisation. The approach of \cite{Roberts2001} involves transforming to a diffusion $\dot{x}_t$ with unit volatility. An alternative scheme is offered by \cite{Chib2006} where the driving Brownian motion of $x_t$ is being used. In these algorithms the ODE states vector $V_{0:n}^{dis}$ becomes a function of $\sigma$, $\dot{x}_{0:n}$ and $\theta_v$. Hence, in a Metropolis step, every proposed value of $\sigma^*$ is associated with the corresponding values of ${V_{0:n}^{dis}}^*$. This succeeds into breaking the perfect dependence between $V_{0:n}^{dis}$ and $\sigma$, even for $m\rightarrow\infty$. But since components of $V_{0:n}^{dis}$ (or functionals thereof) are observed with error, the associated proposed values ${V_{0:n}^{dis}}^*$ should be close to the data for the move to be accepted. As the observation error becomes small and the data increase, this becomes increasingly difficult and leads to very small moves for $\sigma$ and poor MCMC mixing. More details and simulations supporting this argument are provided in the Supplementary Materials (Appendix E). Consequently, we overcome this issue by updating $x_{0:n}^{dis}$ and $\theta$ jointly via the PMCMC algorithm, which is essential as it is not straightforward to implement joint updates with the other approaches mentioned in this section.

\subsection{Adaptive Particle Markov Chain Monte Carlo algorithms}
\label{sec:pMCMC}
Particle filters are SMC algorithms used to recursively explore conditional densities in state space models \citep{Doucet2009}.  For given values of $\theta$, $N$ particles $(\tilde{x}_{i}^{j})$ are sequentially propagated from $t_{0}$ to $t_{n}$. In various time steps $t_{i}$, the trajectories that best fit the data $y_{1:i}$ are given more weight through resampling. Algorithm \ref{alg:SMC} shows how they can be applied in our context.
\begin{algorithm}[h]
\caption{Particle Filter algorithm}
\label{alg:SMC}
\begin{algorithmic}
\STATE \textbf{Initialise:} Set $L^0(\theta)=1$, $W_{0}^{j}=\frac{1}{N}$, sample $(\tilde{x}_{0}^{j})_{j=1,\dots,N}$ from $p(x_0|\theta)$ and calculate $(\tilde{V}_{0}^{j})_{j=1,\dots,N}$ by solving the ODE (for example with the Euler scheme)
\FOR {$i=0$ to $n-1$}
	\FOR {$j=1$ to $N$}
		\STATE Sample $(\tilde{x}_{i:i+1}^{j})$ from \eqref{eq:apprPost} and calculate ($\tilde{V}_{i:i+1}^{j}$) by solving the ODE
		\STATE Set $\alpha^{j}= f(y_{i+1}|\tilde{V}^j_{0:i+1})$
	\ENDFOR
	\STATE Set $W_{i+1}^{j}=\frac{\alpha^{j}}{\sum_{k=1}^{N}\alpha^{k}}$, and $L^{i+1}(\theta)=L^{i}(\theta)\times \frac{1}{N} \sum \alpha^{j}$
	\STATE Resample $(\tilde{V}_{0:i+1}^{j},\tilde{x}_{0:i+1}^{j})_{j=1,\dots,N}$ according to $(W_{i+1}^{j})$,
\ENDFOR		
\end{algorithmic}
\end{algorithm}
The quantity $L^{i+1}(\theta)$ provides unbiased estimates of $p(y_{1:i}| \theta)$ and the resampling step is essential to control the variance of that estimate over time. Algorithm \ref{alg:SMC} also provides a random sample from $p(x_{1:i}| y_{1:n},\theta)$. In order to sample from $\pi(x_{1:n},\theta| y_{1:n})$, the PMCMC algorithm can be used. PMCMC was introduced in \cite{Andrieu2010} and successfully integrates particle filters in MCMC algorithms. Its implementation is presented in Algorithm \ref{alg:PMCMC}.
\begin{algorithm}[h]
\caption{Particle MCMC algorithm (particle Marginal Metropolis Hastings version)}
\label{alg:PMCMC}
\begin{algorithmic}
\STATE \textbf{Initialise:} \; Set current $\theta$ value, $\tilde{\theta}$, to an initial value. Use Particle Smoother (PS) according to Algorithm \ref{alg:SMC} to compute $\hat{p}(y_{1:n}|\tilde{\theta})=L(\tilde{\theta})$ and sample $\tilde{x}_{1:n}^{\tilde{\theta}}$ from $p(x_{1:n}|y_{1:n},\tilde{\theta})$
\FOR {$It=1$ to $NIterations$}
	\STATE Sample $\tilde{\theta}^{*}$ from $Q(\tilde{\theta},.)$
	\STATE Use PS to compute $L(\tilde{\theta}^*)$ and sample $\tilde{x}_{1:n}^{\tilde{\theta}^*}$ from $\hat{p}(x_{1:n}|y_{1:n},\tilde{\theta}^*) $
	\STATE Do $\tilde{\theta}=\tilde{\theta}^{*}$ (and  $\tilde{x}_{1:n}^{\tilde{\theta}}=\tilde{x}_{1:n}^{\tilde{\theta}^{*}}$) with probability $1\wedge\frac{L(\tilde{\theta}^{*})Q(\tilde{\theta}^{*},\tilde{\theta})}{L(\tilde{\theta})Q(\tilde{\theta},\tilde{\theta}^{*})}$
	\STATE Record $\tilde{\theta}$ and $\tilde{x}_{1:n}^{\theta}$
\ENDFOR
\end{algorithmic}
\end{algorithm}
The issues of Section \ref{ssec:gibbs} are now addressed as $x_{0:n}^{dis}$ and $\theta$ are sampled
jointly. In other words $x_{0:n}^{dis}$ is being numerically integrated out, while a sample from its posterior is obtained at each MCMC iteration.

While the PMCMC algorithm is theoretically valid even for a single particle, large values of $N$ are usually required for reasonably stable acceptance rates and large moves in the $\theta$ space; see the Supplementary Materials for a relevant simulation exercise. It is therefore essential to update the $d$-dimensional $\theta$ at once, marking the proposal $Q(\theta,.)$ crucial to the overall MCMC performance. In this paper we propose to use the adaptive Metropolis algorithm of \cite{Roberts2009}. After transforming the parameters to take values in the real line we use a Normal distribution centered at the current value of $\theta$ and with covariance given by $\epsilon \Sigma$. Static random walk metropolis proposals set $\Sigma=I_d$ or $\Sigma=\hat{\Sigma}$ and tune $\epsilon$ to obtain acceptance rate of 0.234. Adaptive schemes change the value $\epsilon$ for each iteration $i$ through diminishing adaptation; e.g. by $\epsilon_{i+1}=\exp\left\{\log(\epsilon_i)  + \alpha_1^n(\text{AccRate}-0.234)\right\}$ where $\alpha_1=0.999$ and `AccRate' denotes the acceptance rate up to iteration $i$. The covariance matrix $\Sigma_{i+1}$ can also be updated as
$$
\alpha_2 \mathcal{N}\left(\theta,\epsilon\frac{2.38^2}{d}\Sigma_0\right) +(1-\alpha_2)\mathcal{N}\left(\theta,\epsilon\frac{2.38^2}{d}\Sigma_i\right)
$$
where $\alpha_2$ is usually set to 0.05, $\Sigma_i$ is the posterior covariance matrix estimated by the draws up to $i$ and $\Sigma_0$ should be specified in advance. In this paper we enhance the above adaptive algorithms utilising information from the EKF to estimate the covariance $\hat{\Sigma}$ or $\Sigma_0$. One choice, EK-Mode, is the observed information matrix at the mode identified by EKF, evaluated through numerical differentiation. Another choice, EK-MCMC, is to run an approximate MCMC scheme based on the EKF approximation of the likelihood and compute the posterior covariance from the draws. Note that the computational burden of these methods is marginal with regards to the PMCMC. As demonstrated in Section \ref{sec:simulations}, the use of EKF can result in substantial improvement.

\section{Simulation Experiments}
\label{sec:simulations}

\begin{figure}[h]
\begin{centering}
\includegraphics[scale=0.3]{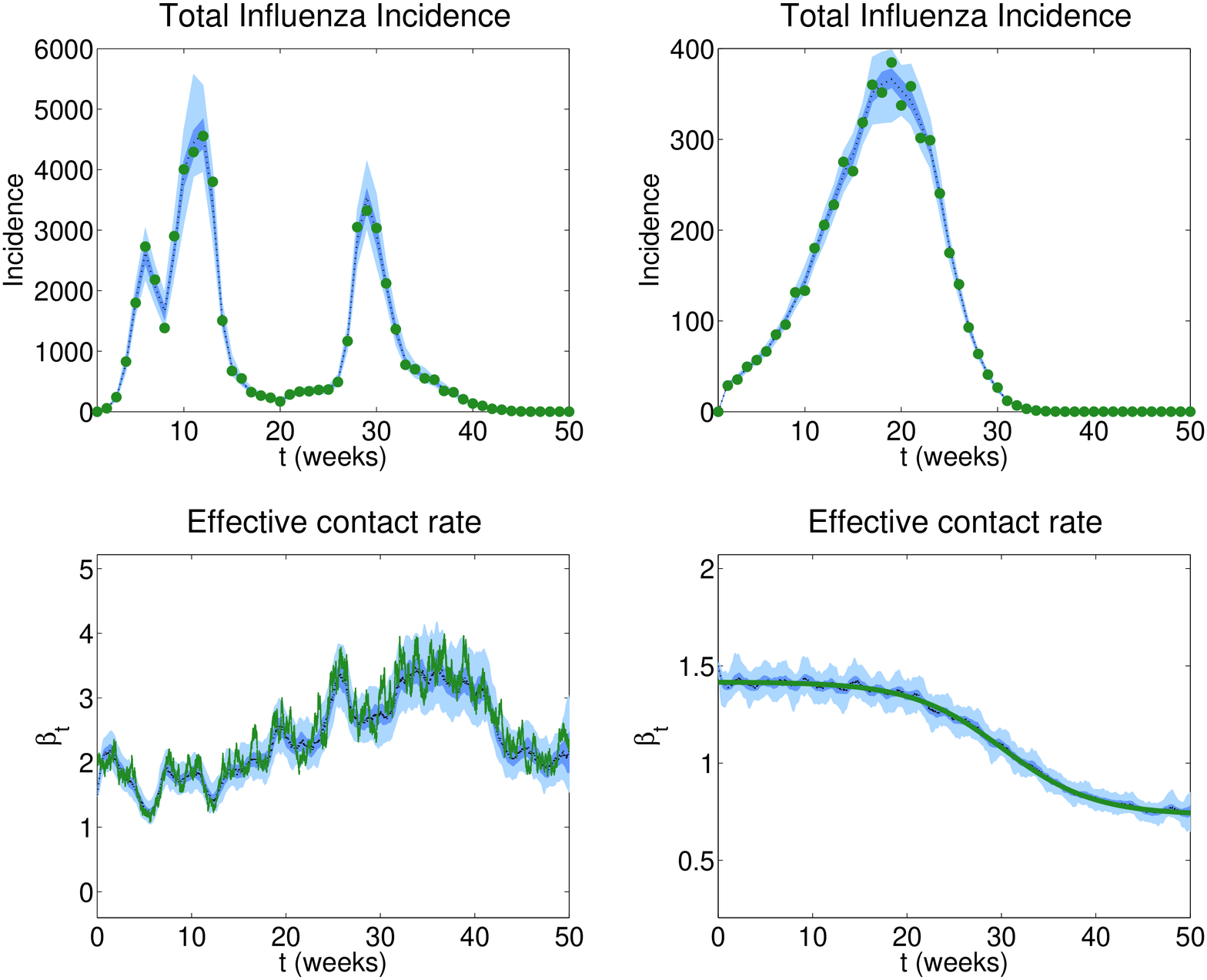}
\par\end{centering}

\caption{\label{fig:AltAndSigm}Illustration of how the underlying dynamic of the effective contact rate can be estimated from weekly recorded cases.\protect \\
Green dots indicate simulated observed incidence (top panels). Green lines indicate simulated effective contact rate trajectories (bottom panels). Black dotted lines indicate the mean of the pointwise posterior density. Dark and light blue areas show credible intervals, respectively at 50\% and 95\% levels.\protect \\
Top panels: simulated weekly numbers of cases observed with noise, and corresponding model-based offline reconstructions (left: experiment 1, right: experiment 2)\protect \\
Bottom panels: simulated and estimated trajectory of the effective contact rate (left: experiment 1, right: experiment 2)}
\end{figure}

The proposed algorithms are illustrated and tested on simulated data in this section. We focus on the BM model, where $\log(\beta_t)$ follows a Brownian motion with volatility $\sigma$, corresponding to the case of little information on the shape of $\beta_t$. The trajectories of $\beta_t$ were drawn either from the BM model itself (experiment 1) or from a deterministic sigmoid curve (experiment 2). The data $y_i$, $i=1,\dots,50$ represent noisy observations of weekly new cases of the epidemic $\int_{week\; i}kE_t dt$. We complete the model by assigning a Normal distribution to each $\log(y_i)$ with mean $\log(\int_{week\; i}kE_t dt)$ and variance $\tau^2$. The parameters were tuned to obtain realistic epidemic incidence curves, and observations were generated setting $\tau=0.1$. The assigned priors were informative for $k$, $\gamma$ and $R(t_0)$ and vague for $E(t_0)$, $I(t_0)$, $\sigma$ and $\tau$, as in Section \ref{ssec:datamodelpriors}. We used 3,000 particles and 100,000 MCMC iterations after a long burn-in period. Fig. \ref{fig:AltAndSigm} shows estimates and 95\% pointwise credible intervals of the path, provided by the adaptive PMCMC initialized with EK-MCMC. The posterior output is in good agreement with the simulation trajectories suggesting that the underlying trajectory of $\beta_t$ can be estimated reasonably well from the partial and noisy observations considered. More can be found in the Supplementary Materials (appendix C), where we also considered a value of $\tau=0.05$ and obtained similar results.

Next, we use the data of experiment 1 to compare the proposed adaptive PMCMC schemes. Comparison is made in terms of the effective sample size $\textrm{ESS}=({1+2\sum_{i\ge 1}\eta(i)})^{-1}$, with $\sum_i\eta(i)$ being the sum of the lagged sample auto-correlations, as in \cite{Geyer1992}. We record the minimum ESS among the MCMC components and multiply by $100$ to monitor the percentage of the total iterations that can be considered as independent. We consider three covariance matrices for each of the two adaptive algorithms defined in Section \ref{sec:pMCMC}: $I_d$ and the ones from EK-Mode and EK-MCMC. For the schemes that adapt $\epsilon$ the minimum ESS was 0.008\% ($I_d$), 0.19\% (EK-Mode) and 0.54\%(EK-MCMC), whereas for the schemes that adapt $\Sigma$ we got 0.57\%, 1.24\% and 1.38\% respectively. Clearly, adapting $\Sigma$ is crucial to obtain a reasonable performance, unless the matrices obtained from EK-Mode or EK-MCMC are used. The proposed adaptive algorithms induce substantial improvement that is expected to intensify as the dimension of $\theta$ increases.

\section{The 2009 A/H1N1 pandemic}
\label{sec:H1N1}
\subsection{Data, model and estimates}
\label{ssec:datamodelpriors}
 The proposed methodology is illustrated on data from the A/H1N1(2009) pandemic in England between June and December 2009. The data consists of estimates of weekly ILI cases $y_{1:n}$ given by the Health Protection Agency \citep{Baguelin2010}. The estimates were obtained from the recorded ILI cases among a selected sample of GPs. They accounted for over-reporting due to similarities in symptoms with other respiratory diseases, based on subsequent virological positivity tests. Corrections for asymptomatic infections and the patients' propensity to consult were also made. Overall the two datasets are different by a multiplicative coefficient $c=10$, whose value is also supported by a further serological survey \citep{Miller2010}. In our analysis $c$ is initially held fixed to $10$, but this choice is explored further in Section \ref{ssec:realtime}. We adopt a model that admits noisy data to reflect the associated uncertainty. The noise model of Section \ref{sec:simulations} was used, combined with a BM formulation of $\mathbb{P}_{x}$. Vague priors, $N_{>0}(0,10^6)$, were put on $\tau$, $\sigma$ and $\beta_0$. The priors for $k$ and $\gamma$ were obtained from additional data sources \citep{Baguelin2010}, the results of which are summarised through Normal distributions that place 95\% probability in a symmetric manner between 1.55 and 1.63 days for the latent period $k^{-1}$, and between 0.93 and 1.23 days for the infectious period $\gamma^{-1}$. A Dirichlet distribution was used for the initial proportions in compartments $S,E,I,R$, constraining the mean of the one in $R$ to be $0.15$, its variance $0.15^2$, and the means of the other initial proportions to be equal.

\begin{figure}[h]
\begin{centering}
\includegraphics[scale=0.3]{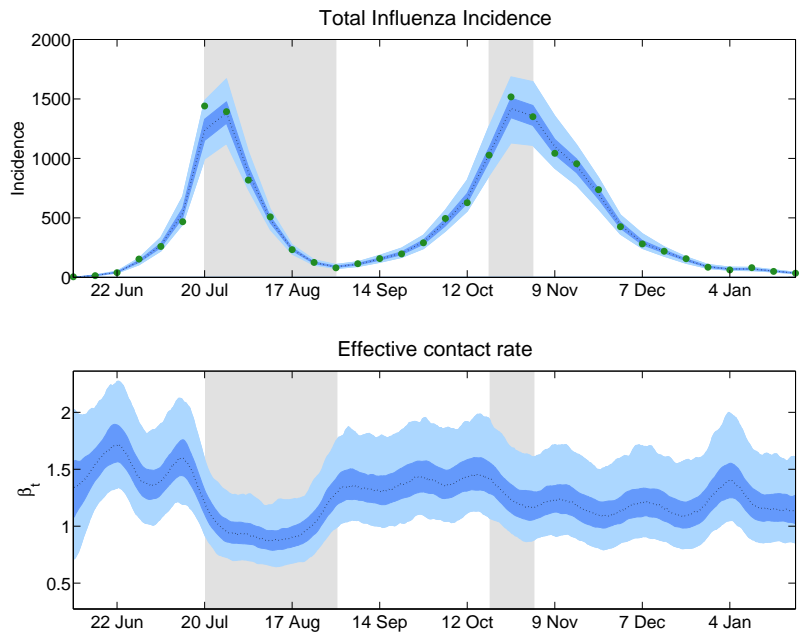}
\par
\end{centering}

\caption{\label{fig:offline}Weekly incidence data from  the A/H1N1 2009  influenza pandemic and corresponding offline estimates of the effective contact rate.\protect \\
Green dots indicate incidence estimates provided by the Health Protection Agency. Black dotted lines indicate the mean of the pointwise posterior density. Dark and light blue areas show credible intervals, respectively at 50\% and 95\% levels.  Holidays are indicated by a light grey area.\protect \\
Top: observations of the weekly total number of  A/H1N1 influenza cases in
London (per 100 000 inhabs.) and model-based offline reconstruction\protect \\
Bottom: {}offline estimates of the effective contact rate. \protect \\ }
\end{figure}

The adaptive EK-MCMC algorithm was applied to the data and Fig. \ref{fig:offline} depicts the incidence curve together with the posterior mean and pointwise 95\% credible intervals. Estimates of $\beta_t$ are also displayed indicating various changes over time. The changes in $\beta_t$ are consistent with the argument that schools closure for holidays have been driving the epidemic: different values are observed during school and holidays periods, appearing to be synchronised with schools opening and closing. Posterior summaries for the static parameters, as well as a sensitivity analysis on the priors can be found in the Supplementary Materials. These suggest that inference is quite sensitive to the choice of prior for $k$ and $\gamma$, but not for the remaining parameters. It would be interesting to repeat the procedure under an evidence synthesis framework and vague priors.

\subsection{Application in real time. Was the first wave waning due to depletion of susceptibles?}
\label{ssec:realtime}
\begin{figure}[h]
\begin{centering}
\includegraphics[scale=0.5]{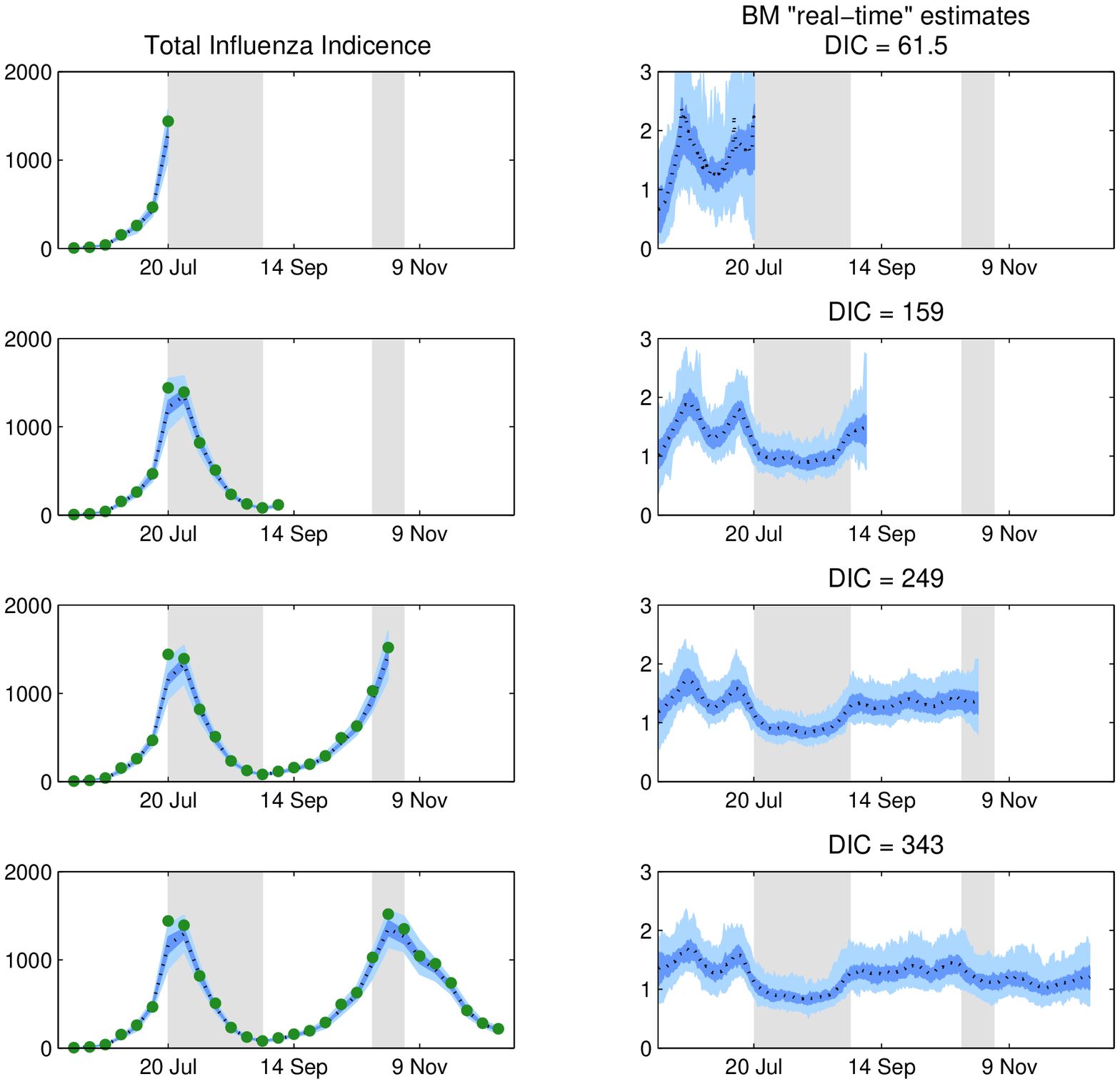}
\par\end{centering}

\caption{\label{fig:What-could-habe}What could have been inferred by carefully
following the epidemic in real time?\protect \\
Green dots indicate observed incidence estimates provided by the Health Protection Agency (left panels). Black dotted lines indicate the mean of the pointwise posterior density. Dark and light blue areas respectively indicate 50\% and 95\% credible intervals of the posterior density. Holidays
are indicated by a light grey area. \protect \\
Left panels: HPA estimates of the weekly total number of  A/H1N1 influenza cases in
London (per 100 000 inhabs.) \protect \\
Right panels: {}``real-time'' estimates of the effective contact rate. \protect \\}
\end{figure}

In this section the methodology of the paper is applied in real time, i.e. considering partial datasets from June 2009 up to the 20th of July, the 7th of September and the 26th of October. Each time the algorithm is run from scratch to provide samples from the joint posterior $\pi(x_{1:i},\theta|y_{1:i})$. From a computational cost point of view this procedure can be improved further by utilising previous MCMC runs, for example under the SMC$^2$ framework \citep{Chopin2011}. We did not pursue this direction further, as the PMCMC algorithm runs quite fast (less than 2 hours on a standard PC). In order to reduce uncertainty, especially at early stages, the value of $\tau$ was set to $0.1$ rather than being estimated as in Section \ref{ssec:datamodelpriors}. We otherwise use the same model as before. A model with integrated Brownian motion was also fit but BM was chosen in terms of DIC; see Supplementary Materials (Appendix C). The main results are shown in Fig. \ref{fig:What-could-habe}.

On August $1^{st}$, the first wave of the epidemic had waned, incidence rates were decreasing and schools had closed. There were two competing scenarios to explain the epidemic decline: (i) holidays had caused the waning of the epidemic by lowering the effective contact rate. Hence, a similar or stronger wave could occur when schools would reopen in September in colder climatic conditions. (ii) The epidemic had stopped independently of holidays because a critical proportion of the population had been infected, conferring a sufficient level of herd immunity to stop the epidemic. In this case, no second wave was to be expected in September. On August $1^{st}$ there was great uncertainty around the value of c \citep{Baguelin2010}, which is crucial in distinguishing between the two scenarios. We therefore conducted the following exercise.

\begin{figure}[h]
\begin{centering}
\includegraphics[scale=0.5]{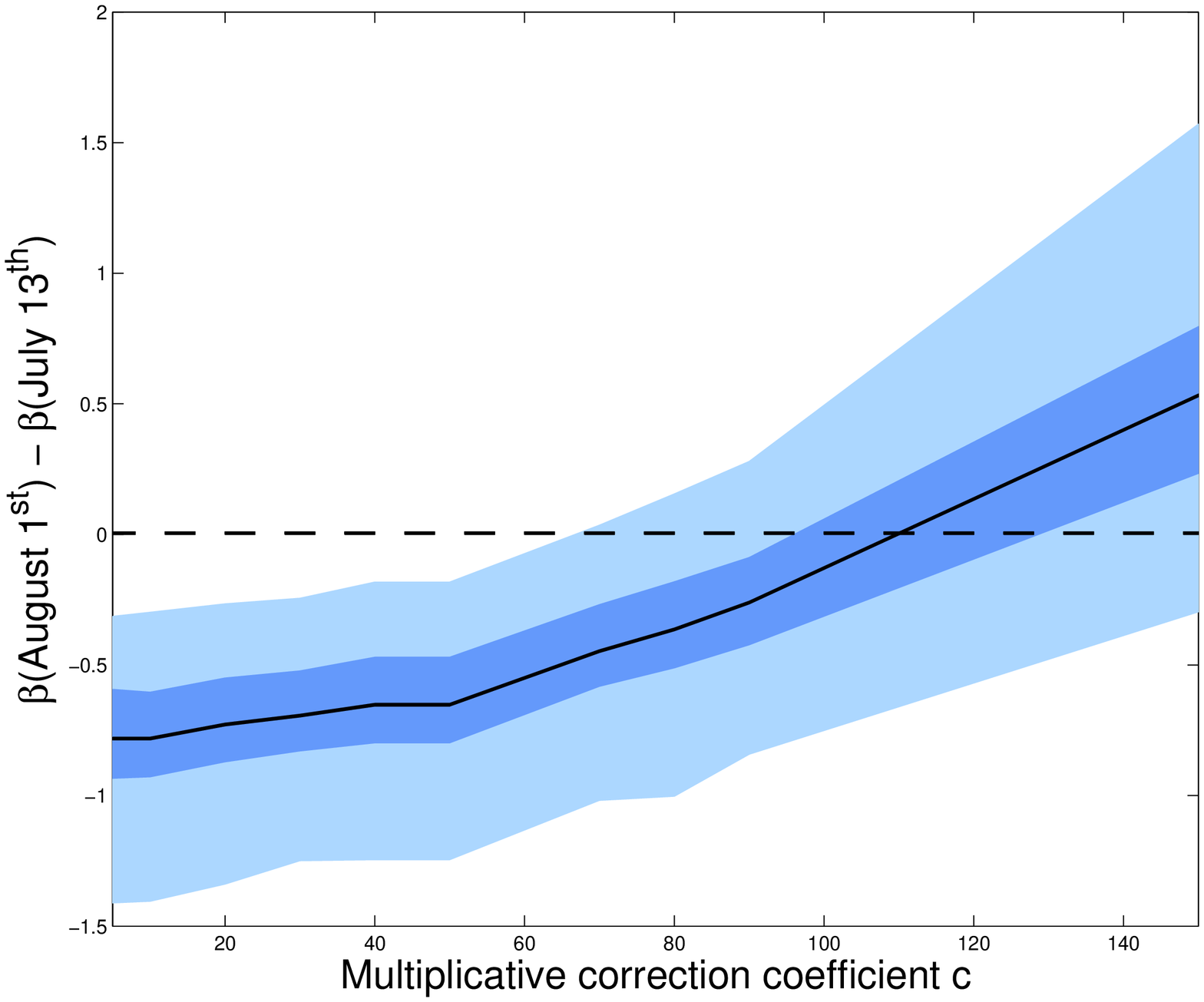}
\par
\end{centering}
\caption{\label{fig:The-implication-of}The implication of different scenarios
for the real value of underreporting on the decrease of the effective contact rate
between July $13^{th}$ and August $1^{st}$ \protect \\
For each value of c, the mean of the posterior density for $\beta(August\;1^{st})-\beta(July\;13^{th})$ is plotted in black. Dark and light blue areas respectively indicate 50\% and 95\% credible intervals of the posterior density. The dotted line locates the scenario whith no change in the effective contact rate.}
\end{figure}

The PMCMC algorithm, run up to August 1st, provides samples from the posterior of the difference in $\beta_t$ between July $13^{th}$ (before the decrease in incidence) and August $1^{st}$. For $c=10$, the $97.5\%$ point of this posterior  is $-0.32$, indicating a decrease in $\beta_t$. The latter supports scenario (i), as the competing scenario is associated with a zero-decrease in $\beta_{t}$. Nevertheless, as this value depends on $c$, the algorithm was run for different values of it ranging from 20 to 150. The results appear on Fig. \ref{fig:The-implication-of}. Note that the $97.5\%$ point of interest increases as a function of $c$ and reaches $0$ for a correction factor close to $70$. As this level seemed unrealistic \citep{Baguelin2010}, the experiment provides evidence in favour of scenario (i) highlighting the danger of a second wave in September, that actually occurred. Such evidence can be important for decision-makers, especially when considering implementations of preventive measures as vaccines.

\subsection{A multiple age group diffusion driven SEIR model}
\label{ssec:multiplegroups}

\begin{figure}[h]
\begin{centering}
\includegraphics[scale=0.5]{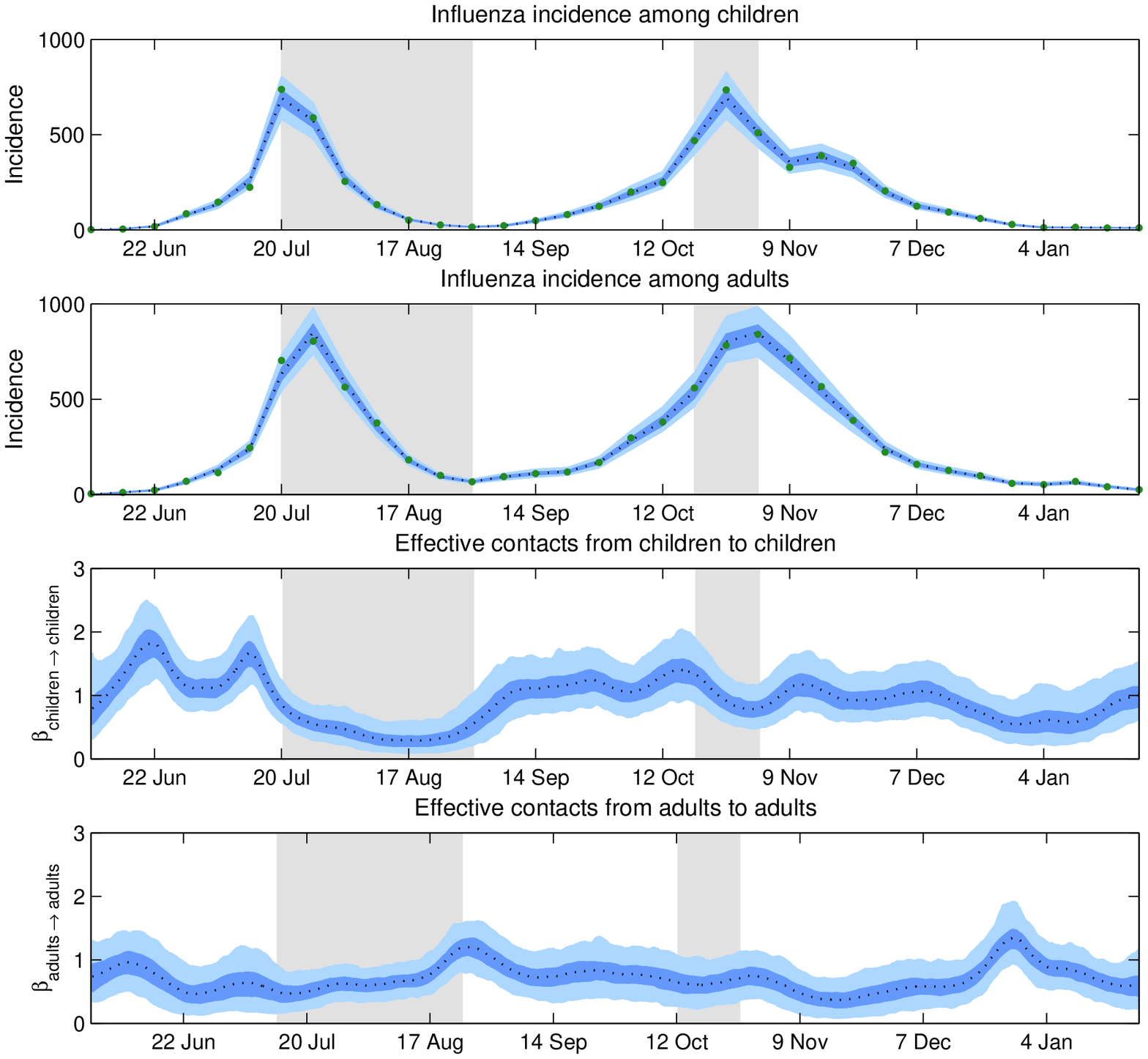}
\par\end{centering}

\caption{\label{fig:multiplegroups} Offline estimates of the effective contact rate among children and adults during the A/H1N1  2009 influenza pandemic using a 2-classes age-structured model and age-specific incidence data.\protect \\
Green dots indicate observed incidence estimates among each age group provided by the Health Protection Agency (first and second panels). Black dotted lines indicate the mean of the pointwise posterior density. Dark and light blue areas respectively indicate 50\% and 95\% credible intervals of the posterior density. Holidays
are indicated by a light grey area. \protect \\
First panel: HPA estimates of the weekly total number of A/H1N1 influenza cases among children in London (per 100 000 inhabs.)\protect \\
Second panel: HPA estimates of the weekly total number of  A/H1N1 influenza cases among adults in London (per 100 000 inhabs.) \protect \\
Third panel: {}offline estimates of the effective contact rate from children to children. \protect \\
Fourth panel: {}offline estimates of the effective contact rate from adults to adults. \protect \\
}
\end{figure}

The analysis of Section \ref{ssec:datamodelpriors} can be used to construct more structured models. For example, the effect of holidays is evident and may differ from children to adults, thus casting doubts on the assumption of a homogeneous population. It seems more natural to consider a model with two age groups (c:children and a:adults) and target all possible effective contact rates among them. In our notation $\beta^{ca}$ refers to the effective contact rate from children to adults and $S^c$ denotes the number of susceptible children. For reasons of parsimony we assign Brownian motions to $\log(\beta^{cc}_t)$, $\log(\beta^{aa}_t)$ and treat $\beta^{ca}$, $\beta^{ac}$ as constant. We also set we set $\beta^{ca}=\beta^{ac}=b$, in line various multiple age groups epidemic models in different settings (e.g. \cite{Whitaker2004}). The dynamic part of the model is now given by
\begin{equation}
\begin{cases}
\frac{dS^{c}_{t}}{dt}=-S_{t}^{c}\left(\beta_{t}^{cc}\frac{I_{t}^c}{N^c}+b\frac{I_{t}^a}{N^a}\right),\;\; \frac{dE_{t}^{c}}{dt} =S_{t}^{c}\left(\beta_{t}^{cc}\frac{I_{t}^c}{N^c}+b\frac{I_{t}^a}{N^a}\right)-kE_{t}^c,\\
\frac{dS^{a}_{t}}{dt}=-S_{t}^{a}\left(\beta_{t}^{aa}\frac{I_{t}^a}{N^a}+b\frac{I_{t}^c}{N^c}\right),\;\; \frac{dE_{t}^{a}}{dt} =S_{t}^{a}\left(\beta_{t}^{aa}\frac{I_{t}^a}{N^a}+b\frac{I_{t}^c}{N^c}\right)-kE_{t}^a,\\
\frac{dI_{t}^c}{dt} =kE_{t}^c-\gamma I_{t}^c, \;\; \frac{dR_{t}^c}{dt} =\gamma I_{t}^c,\;\;\frac{dI_{t}^a}{dt} =kE_{t}^a-\gamma I_{t}^a, \;\; \frac{dR_{t}^a}{dt} =\gamma I_{t}^a.
\end{cases}\label{eq:multipleSEIR}
\end{equation}
The data from the A/H1N1(2009) pandemic provide incidence estimates for  children and adults separately so they can be used to estimate the model of \eqref{eq:multipleSEIR}. If only final outcome data were available, not all effective contact rate parameters would be estimable. However, the temporal dataset provides extra information by the relative variation of susceptible and infective population in adults versus children. We applied the EK-MCMC scheme, which was essential in order to obtain reasonable MCMC performance. Fig. \ref{fig:multiplegroups} depicts the results. Unlike earlier attempts with versions of a multi-group model with a single diffusion driving all contact rates, the fit appears to be good. The trajectory of children seems to be similar with that of Fig. \ref{fig:offline} thus stressing their role to the evolution of the epidemic. More details, including posterior summaries for the parameters and information about the priors can be found in the Supplementary Materials (Appendix C).

\section{Discussion}
In this paper we examined epidemic models where some of the parameters are represented by diffusions
or integrals thereof. The main motivation was to account for various time varying drivers
(virus evolution, seasonality, schools closure, etc), while maintaining a simple interpretation.
We present a unified framework that supports data augmentation MCMC schemes based
on fine partitions on the diffusion path. The associated approximation error can be controlled
by the user without affecting the MCMC performance and can be viewed as an extension of the
approaches by \cite{Roberts2001,Chib2006} to the more challenging observation regime of this paper. The consideration of the algorithms in a continuous time setting
revealed major issues associated with Gibbs data-augmentation schemes. This justifies
the use of particle MCMC, which updates paths and parameters jointly, while pointing directions
for future research on Gibbs schemes. We
also presented a computational machinery based on the PMCMC algorithm \citep{Andrieu2010}, that was integrated in an adaptive MCMC context. We consider EKF based adaptive algorithms
that can offer substantial improvement, especially in cases with many static parameters. This paper is one of the first applications of PMCMC in epidemic
models and data; standard PMCMC schemes were also used in \cite{Rasmussen2011}.

Initially we relied on a simple SEIR model but such an analysis can be viewed as an exploratory
tool towards more structured models; e.g. the age-structured model of Section 5.3 that appears
to be an improved representation of reality. This approach can help in developing richer models and
testing alternative scenarios for public health interventions, or to bring further insights on extrinsic
factors such as climate on the dynamics of epidemics. Moreover, this framework
can support multiple sources of data, of potentially different nature: \cite{Rasmussen2011} has shown how time series and genealogies can be combined in a PMCMC inference framework for more informative estimates. While we worked mainly with influenza time series, the developed methodology can be applied to other cases; current work considers its application as part of the CHARME project \citep{Boily2007}. The presented approach may also be thought as an alternative to the white noise modeling of environmental stochasticity introduced in \cite{Breto2009}, as it offers to the possibility to capture the dynamics of environmental drivers. A potential next step will be to combine environmental with demographic stochasticity, modelling infections as Poisson processes which rates depend on a time-varying $\beta$.

The inferential framework presented in this article shares the "plug and play" feature of the Iterated Filtering methodology. While extra care and further study is required for specific models or datasets, its algorithmic aspects can be decoupled from the modeling aspects. This provides the possibility to develop generic inference packages: we are currently working towards its integration in a generic inference platform inspired from the R package POMP.

\section{Supplementary Materials}
Supplementary material is available online at http://biostatistics.oxfordjournals.org. It contains implementation details for the PMCMC algorithm (Appendix A), a comparison with the EKF (Appendix B), additional information on Sections \ref{sec:simulations} and \ref{sec:H1N1} (Appendix C). Also the sensitivity analysis (Appendix D) and a detailed exposition of the issues of Section \ref{ssec:gibbs} (Appendix E).

\section*{Acknowledgments}
The authors would like to thank John Edmunds, Nikolaos Demiris, Wicher Bergsma and the three anonymous reviewers for their helpful and constructive comments.

\bibliographystyle{biorefs}
\bibliography{Biblio}

\newpage

\end{document}


\title{Capturing the time-varying drivers of an epidemic using stochastic dynamical systems \\ SUPPLEMENTARY MATERIALS}

\author{JOSEPH {DUREAU}$^\ast$, KONSTANTINOS {KALOGEROPOULOS}\\[4.1pt]
{\it Department of Statistics , London School of Economics}\\
{\it and Political Science, UK{}}\\[2pt]
{j.dureau@lse.ac.uk}\\
MARC  {BAGUELIN}\\[4.1pt]
{\it Immunisation, Hepatitis and Blood Safety Department,} \\
{\it Health Protection Agency, }\\
{\it Centre for the Mathematical Modeling of Infectious} \\ 
{\it  Diseases, London School of Hygiene and Tropical Medicine}}

\maketitle

\pagestyle{headings}

\markboth{J. Dureau, K. Kalogeropoulos and M. Baguelin}{Capturing the time-varying drivers of an epidemic}

\maketitle

\footnotetext{To whom correspondence should be addressed.}

Here we provide supplementary information for various parts of the main paper. Appendix A illustrates the effect of various algorithmic parameters and suggestions on how they can be set in practice. Appendix B presents a simulation based comparison of the particle filter with the Extended Kalman Filter (EKF), whereas in Appendix C we provide more details on the analyses of Sections 4 and 5 of the main paper. Appendix D contains a sensitivity analysis for the priors assigned to the parameters of the model in Section 5.1. Finally, in appendix E, more details are given regarding the formulation of the models in continuous time and their potential implications on the associated data augmentation schemes.

\section*{\label{ApenA}Appendix A: details of the PMCMC implementation}

In this Appendix, we provide more details for the practical implementation of the PMCMC algorithm presented in this article. We specify how to determine key parameters of the algorithm, i.e. the Euler discretisation time-step $\delta$, the number of particles $N_{parts}$ used in the Particle Smoother (PS), and how to set the Metropolis updates of the parameter vector $\theta$.

\subsection*{Determining the Euler discretization time-step}
\begin{figure}[h]
\begin{centering}
\includegraphics[scale=0.4]{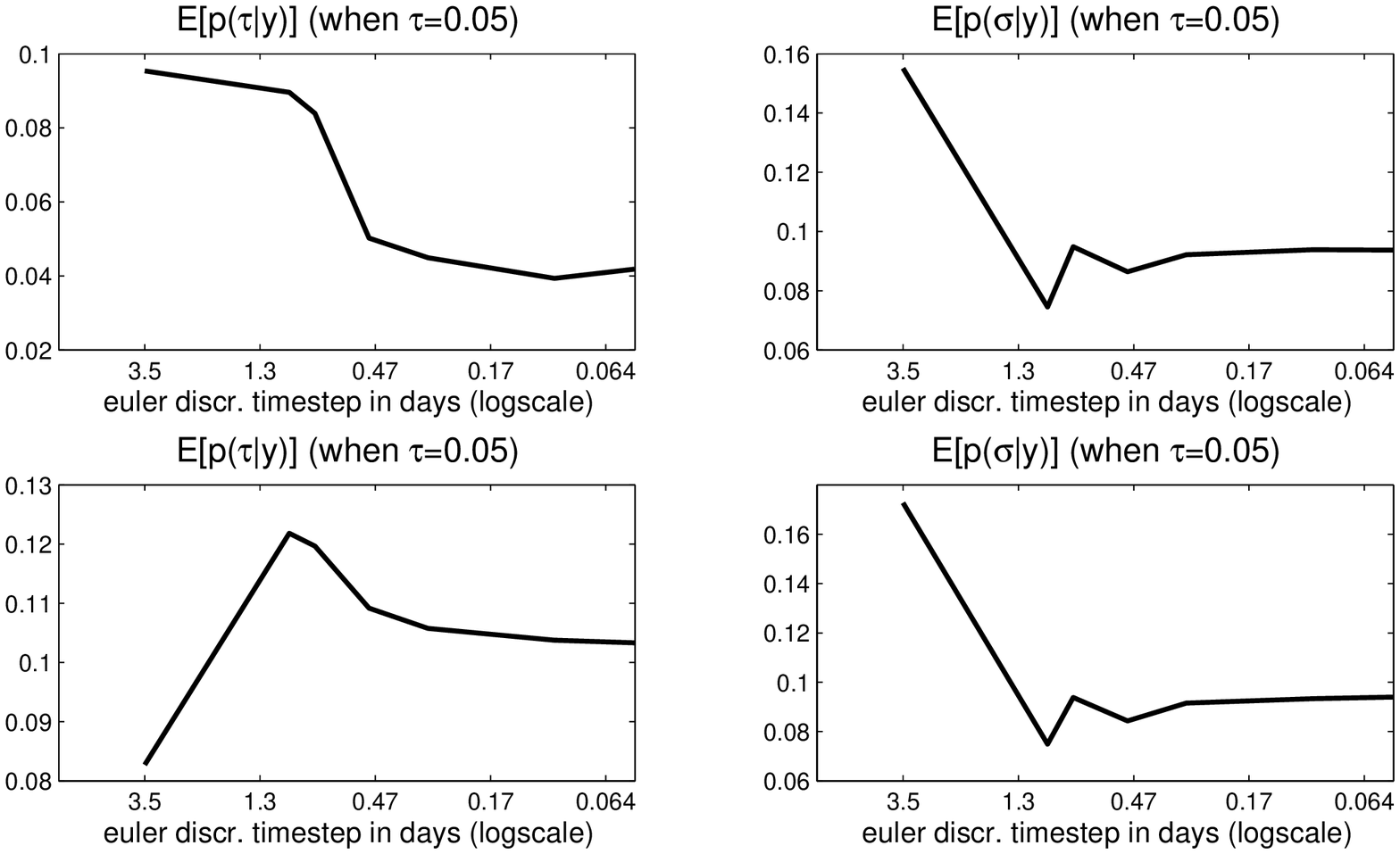}
\par\end{centering}
\caption{\label{fig:Euler}Convergence of the posterior density as the Euler discretization time-step $\delta$ decreases (x-axis in the log-scale)}
\end{figure}

In general, solutions of the nonlinear ODEs encountered in epidemic models are not available in closed form. In order to evaluate $\pi(x_{0:n},\theta|y_{1:n})$, the trajectory of the system needs to be discretised according to a given time-step $\delta$ to provide an approximate solution. The Euler approximation ensures that as $\delta$ tends to 0, $\hat{\pi}_{\delta}(x_{0:n},\theta|y_{1:n})$ converges to $\pi(x_{0:n},\theta|y_{1:n})$. In practice a sequence of decreasing $\delta$ values is chosen and quantities such as $E[\hat{p}_{\delta}(\sigma|y_{1:n})]$ or $E[\hat{p}_{\delta}(\tau|y_{1:n})]$, are monitored. For sufficiently small values of $\delta$, a convergence is generally observed, as shown in Fig. \ref{fig:Euler} for two different datasets of weekly influenza data. In this case, a $\delta=0.1$ day would be a reasonable choice. We note at this point that the computational cost is of $O(\delta^{-1})$

\subsection*{Determining an optimal number of particles}
\begin{figure}[h]
\begin{centering}
\includegraphics[scale=0.4]{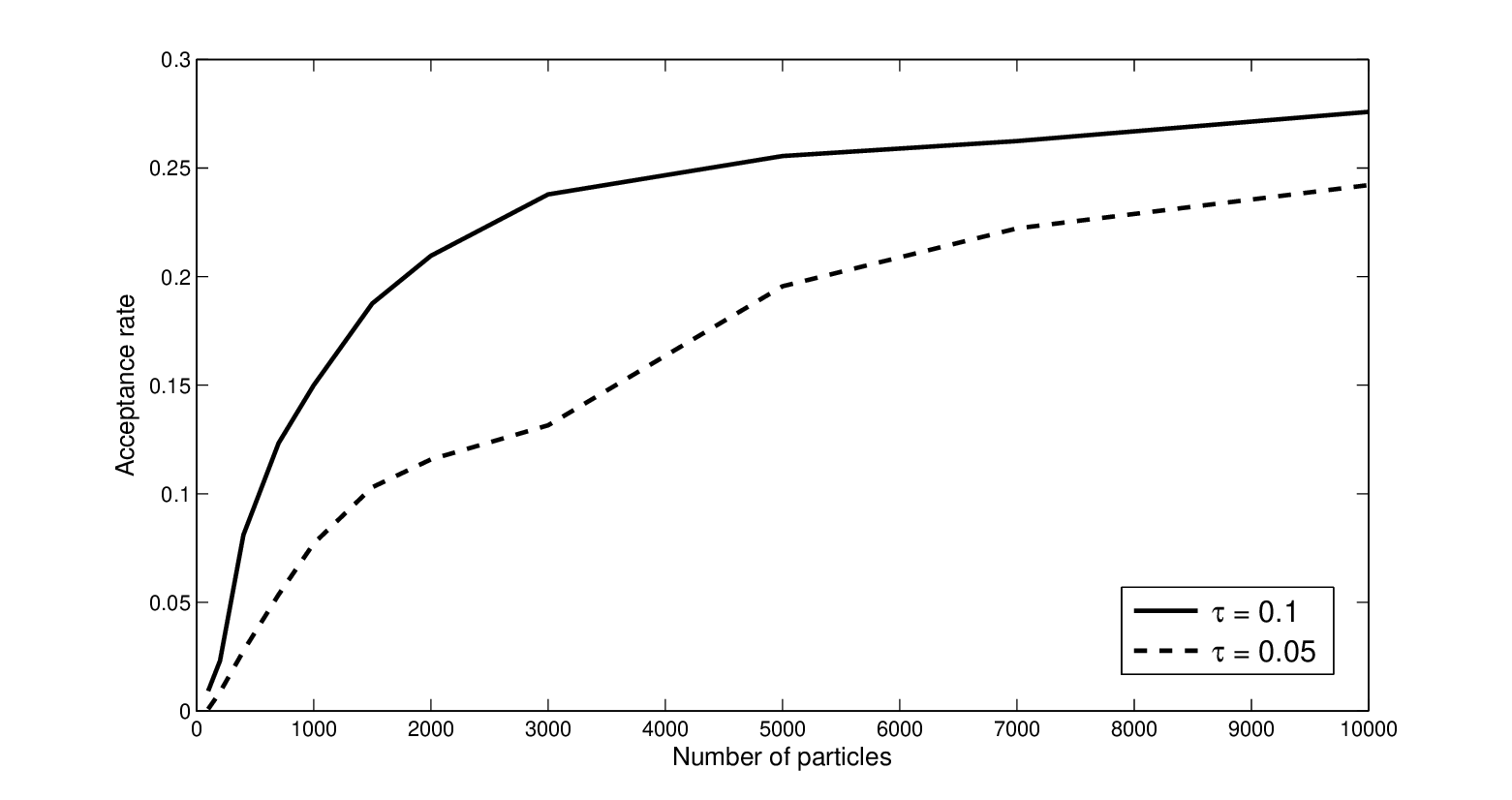}
\par\end{centering}
\caption{\label{fig:NbParts}Acceptance rate as a function of $N_{parts}$, in two situations where the noise amplitude is respectively $10\%$ (full line) and $5\%$ (dotted line).  }
\end{figure}

The PMCMC algorithm is theoretically valid regardless of the number of particles $N_{parts}$ used in the particle smoother, as shown in \cite{Andrieu2010}. Nevertheless, the smaller the number of particles used in the particle smoother, the more noisy the estimate of the likelihood $\hat{p}(y_{1:n}|\theta)$ becomes. This noise has a negative impact on the acceptance rate of the MCMC algorithm run in the $\theta$ space. Consequently, $N_{parts}$ has to be big enough so that it won't affect the acceptance rate, keeping in mind that the cost is of $O(N_{parts})$. Hence, a compromise needs to be achieved. Fig. \ref{fig:NbParts} shows how the acceptance rate increases as the number of particles gets higher. Note that the acceptance rate has a plateau form, indicating that it is perhaps not worthwhile to increase $N_{parts}$ beyond some point. We repeat the experiment for two different values of the measurement error parameter ($\tau$). This figure shows that when the observational noise decreases, more particles are needed, which is explained by the fact that the particles need to be `closer' to the observations.

\section*{\label{ApenB}Appendix B: Assessing the validity and limitations of the Extended Kalman approximation}

Here, we compare the particle filter with the EKF approach to assess the the extent of the gain of avoiding the approximations of the latter. A set of 100, 7-month long, time-series of weekly influenza cases were drawn from the BM model. In order to ensure realistic epidemic datasets, we `reverse-engineered' randomly selected influenza time-series $(y_{1:n}^{Goog,j})_{j=1}^{100}$ from the freely available Google FluTrend data \citep{Ginsberg2008}. For each of the datasets, we obtained estimates of $(\beta_{0:n}^{Sim,j})_{j=1}^{100}$ and the corresponding parameters $(\theta_{j})_{j=1}^{100}$. These quantities were then used to generate influenza time-series $(y_{1:n}^{Sim,j})_{j=1}^{100}$. The static parameters of the model (initial conditions, $k$, $\gamma$, $\sigma$ and $\tau$) were assumed known, to isolate the problem of estimating $\beta_{t}$ from accounting for parameter uncertainty and perform more relevant comparisons. We compare the following two estimators of $\beta_{i}^{Sim,j}$: $\hat{\beta}_{i}^{Filt,j}= \hat{E}(\beta_{i}^{j}|y_{1:i}^{j},\theta_{j}^{*})$ obtained from the filtering distribution where $\hat{E}(.)$ denotes Monte Carlo estimates of the relevant expectations, and the the EKF estimator $\hat{\beta}_{i}^{EKF,j}= \tilde{E}^{EKF}(\beta_{i}^{j}|y_{1:i}^{j},\theta_{j}^{*})$ where $\tilde{E}^{EKF}(.)$ denotes expectation under EKF. The performance of the estimators is measured through their bias and Mean Squared Error (MSE). The results indicate a better performance for $\hat{\beta}_{i}^{Filt,j}$. The bias of the estimates provided by the EKF is $0.0285$ while use of $\hat{\beta}_{i}^{Filt,j}$ reduces the bias by about $78\%$ (0.0063). The corresponding relative reduction in MSE is smaller ($10$\%, $0.0270$ to $0.0242$), indicating a bias-variance tradeoff. Use of the smoothing distribution estimator $\hat{\beta}_{i}^{Sm,j}= \hat{E}(\beta_{i}^{j}|y_{1:n}^{Sim,j},\theta_{j}^{*})$ is associated with a further $87$\% ($0.0032$) reduction in the MSE, while keeping the bias at the same low levels. The estimators $\hat{\beta}_{i}^{Filt,j}$ and $\hat{\beta}_{i}^{Sm,j}$ are associated with a tolerable computational cost of 2 hours on a standard PC.
In conclusion, the bias introduced by the Extended Kalman approximation is non-negligible with regards to the level of accuracy that can be obtained with exact particle methods on this type of datasets. Nevertheless, this study has shown the approximation to be robust and motivates the use of the approximated model as a proxy for the exact one, for example to initialize the particle MCMC algorithm in the way presented in this article.

\section*{\label{ApenC}Appendix C: details on the simulations and results of sections 4 and 5}

\subsection*{Assessing the validity of the particle MCMC through simulations}
This section deals with a series of experiments that were conducted in order to assess the validity of the particle MCMC, and to illustrate on different examples how the trajectory $\beta_{1:n}$ could be captured from noisy weekly cases observations. Experiment 1 is based on a trajectory of $\beta_{1:n}$ simulated from a random walk model with volatility $\sigma^2=0.07^2$. Two different corresponding epidemic datasets have been generated, for given and equal initial conditions and biological parameters, respectively with observational noise $\tau=0.1$ (experiment 1.a) and $\tau=0.05$ (experiment 1.b). Similarly, two epidemic datasets resulting from an effective contact rate following a significantly decreasing sigmoid were generated with respectively $\tau=0.1$ (experiment 2.a) and $\tau=0.05$ (experiment 2.b).

\begin{figure}[h]
\begin{centering}
\includegraphics[scale=0.5]{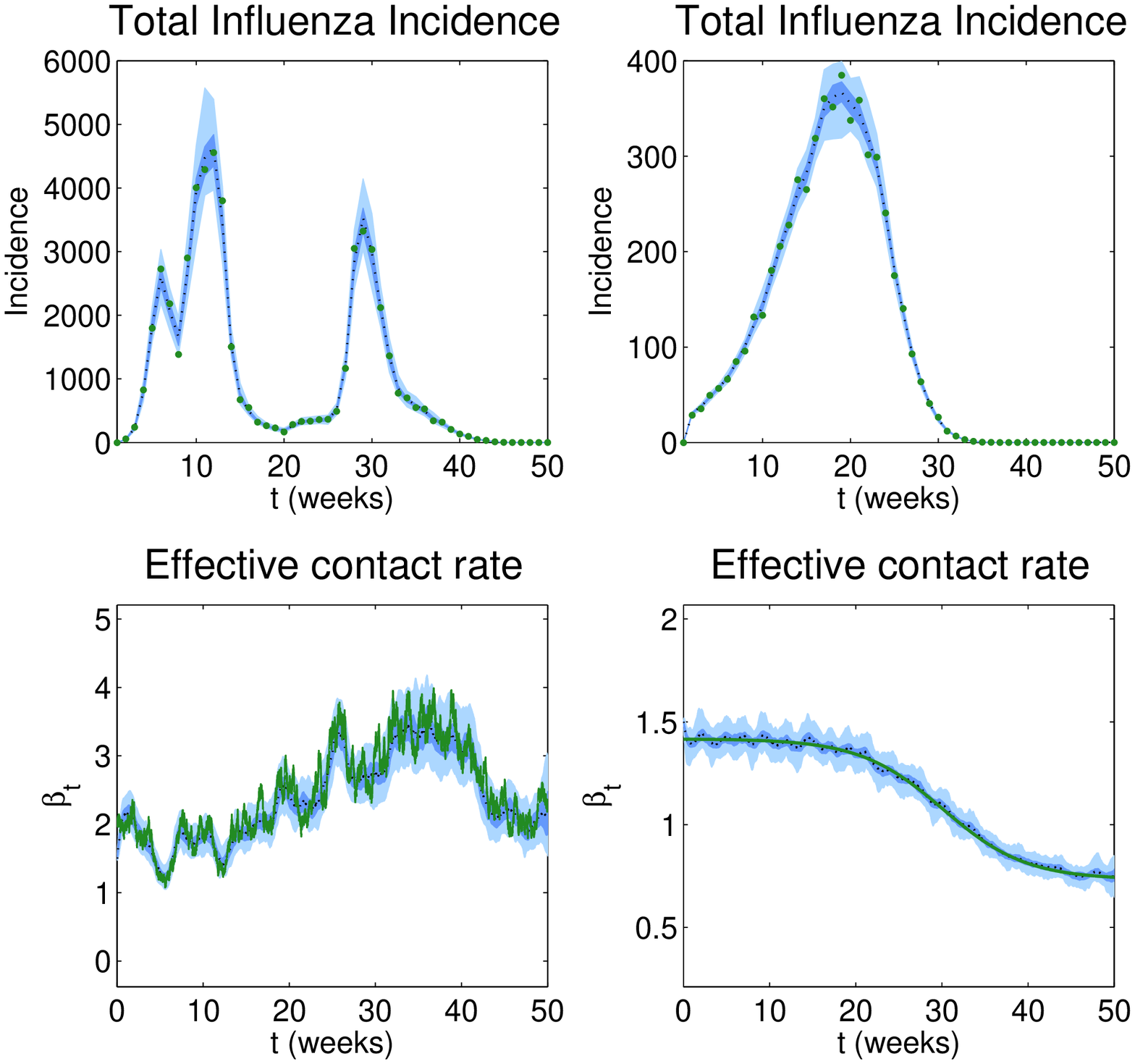}
\par\end{centering}

\caption{\label{fig:AltAndSigm}Illustration of how the underlying dynamic of the effective contact rate can be estimated from weekly recorded cases ($\tau=0.05$)\protect \\
Green dots indicate simulated observed incidence (top panels). Green lines indicate simulated effective contact rate trajectories (bottom panels). Black dotted lines indicate the mean of the pointwise posterior density. Dark and light blue show credible intervals, respectively at 50\% and 95\% levels.\protect \\
Top left: experiment 1.b, weekly number of cases observed with noise\protect \\
Top right: experiment 2.b, weekly number of cases observed with noise\protect \\
Bottom left: experiment 1.b, simulated and estimated trajectory of the effective contact rate\protect \\
Bottom right: experiment 2.b, simulated and estimated trajectory of the effective contact rate
}
\end{figure}

For each of these datasets, our proposed methodology was run to estimate $\sigma$, $\tau$ and $\beta_{1:n}$. cFig. 1 of the main text contains estimates of the incidence time series as well as $\beta_{1:n}$ for the experiments with $\tau=0.1$ (Exps 1.a and 2.a). Corresponding figures for Experiments 1.b and 2.b are shown in Fig. \ref{fig:AltAndSigm} of this document. Moreover, Table \ref{table:TauSigmaEsts} presents the mean, median and $95\%$ credible intervals for the estimates of $\sigma$ and $\tau$ in each of the experiments. In experiment 2 the estimates seem to be in good agreement with the true values, as the latter are contained in the 95\% credible intervals. The aim was to assess the robustness of the proposed methodology to model mis-specification, fitting a Brownian motion to a smooth sigmoid curve. The algorithm performs reasonably well, succeeding in capturing the trajectories of $\beta_t$ and all the parameters except for $\tau$ which is slightly underestimated. A potential explanation for this is that part of the variability is absorbed from the volatility parameter of the Brownian motion.

\subsection*{A/H1N1 pandemic}
We provide here with the corresponding trace plots (Fig. \ref{fig:Trace}) for the parameter estimates of Section 5.2. The trace plots indicate good mixing of the PMCMC algorithm for every parameter. This was achieved by following the procedure presented in Appendix A.

\subsection*{Age-structured model: children and adults}
Additional information on the analysis of section 5.3. can be found in Table  \ref{table:StructEsts}.

\subsection*{\label{ApenE}Illustration of alternative approaches on the real time example}
In Fig. \ref{fig:Comparing}, we repeat the real-time analysis, conducted in the Section 5.2 of the main paper, under two alternative approaches. First, we consider a model with an integrated Brownian motion (iBM) on $x_t$, implying smoother $\beta_t$ trajectories as opposed to the non-differentiable paths induced by the Brownian motion (BM) formulation. The choice between those models is not trivial and could depend on the context of the epidemic. We decided to adopt the model with Brownian motion (BM) on the basis of the Deviance Information Criterion (DIC) of \cite{Spiegelhalter2002}; as we can see from Fig. 3 of the main document and Fig. \ref{fig:Comparing} of the SM, the BM model is consistently better in that respect. Nevertheless, there is uncertainty on the performance of the DIC criterion in the setting of this. It would be quite interesting to explore this further in the future and compare with alternative model choice criteria.

The second approach adopts and applies the methodology of maximum likelihood via iterated filtering (MIF), introduced in \citep{Ionides2006}. Initially, estimates $\theta$ are obtained by maximising $p(y_{i:n}|\theta)$ subject to some constraints set by the priors. Second, a particle filter was run with $\theta$ fixed to its estimated value. As expected, given that we are not accounting for parameter uncertainty, the resulting pointwise 95\% credible intervals are narrower; roughly 50\% on the 6-month dataset and even more at early stages with less information on $\theta$.

\section*{\label{ApenD}Appendix D: Sensitivity analysis}
We explore in this section the robustness of the obtained estimates. We concentrate on the example of Section 5.1 in the main paper, where the observational noise was estimated and informative priors were used for the initial proportion of immune individuals in the population ($R(0)$), and the lengths of the latent ($k^{-1}$) and infectious periods ($\gamma^{-1}$ ). For each of these quantities, the mean of the prior densities have been tilted by respectively -20\%, -10\%, +10\% and +20\%. In all cases but for the tilted parameters themselves, the median of the corresponding posterior densities lie both in the 95\% and 50\% credible intervals of the untitled case. The resulting medians estimates are shown in table \ref{table:Sensitivity}, along with the originally obtained summary statistics. Furthermore, for all the runs with tilted priors, the resulting $\beta_{1:n}$ trajectories remain completely in the original 50\% credible intervals.

\section*{\label{ApenF}Appendix E: details of Gibbs data augmentation scheme}

In this section we provide more details on the Gibbs schemes discussed in section 3.3 of the main text. Stochastic epidemic models presented in this paper can be written as
\begin{equation}
\begin{cases}
dx_{t} & =\mu_{x}(x_{t},\theta_x)dt+\sigma_{x}(x_t,\theta_x)dB_{t},\;\;0<t<t_n\\
y_{1:n}| V_{0:n},\theta_y &\sim \mathbb{P}_{y}(y_{1:n}|V_{0:n},\theta_y),\;\;\;\;V_{0:n} =g(x_{0:n},\theta_v)
\end{cases}\label{eq:PartObsDiffusion}
\end{equation}
\noindent where $V_t$ represents the ODE states vector observed trough partial and noisy data $y_{1:n}$. The rest of the model is defined in section 3.1. Since it contains intractable densities we work with the time discretised versions $x_{0:n}^{dis}$ and $V_{0:n}^{dis}$ and proceed using the Euler approximating scheme. A Gibbs algorithm alternates between updating the trajectories of $x_{0:n}^{dis}$ (and consequently $V_{0:n}^{dis}$) given $\theta$, and vice versa. Nevertheless, as the Euler time step $\delta$ goes to 0, the quadratic variation process of $x_t$ uniquely determines the value of $\theta_x$ in $\sigma_x(.)$ and the algorithm degenerates \citep{Roberts2001}. In practice this translates into a mixing time of $O(m)$.

In the context of diffusion driven epidemic models this problem was dealt with suitable reparametrisations such as the ones in \cite{Chib2006} or \cite{Kalogeropoulos2007}. The latter uses the Lamperti transform, i.e.
$x_t\rightarrow H(x_t,\theta_x)=\eta(x_t,\theta_x)-\eta(x_0,\theta_x)=:u_t$ where $\eta(\cdot;\theta_x)$ is an antiderivative
of $\sigma_x^{-1}(\cdot;\theta_x)$. Assuming that $\sigma_x(\cdot;\theta_x)$ is continuously
differentiable, an application of Ito's lemma provides the SDE of the transformed diffusion $u_t$ as:
\begin{equation}
  \label{eq:diffusionunitvol}
  du_t=\nu(u_t;\theta_x)dt+dB_t\  ,\quad  u_0=0\ ,
\end{equation}
where
\begin{equation*}
\nu(u_t;\theta_x)=\frac{\mu_x\big(H^{-1}(u_t,\theta_x),\theta_x\big)}{\sigma_x\big(H^{-1}(u_t,\theta_x),\theta_x\big)}- \frac{1}{2} \sigma_x^{\prime}\big(H^{-1}(u_t,\theta_x),\theta_x\big)\ .
\end{equation*}
Let $\mathbb{P}^{u}$ denote the distribution of $u_t$. Girsanov formula provide its density with respect to that of a standard Brownian motion, denoted by $\mathbb{W}$,
\begin{equation}
  \label{eq:likeunitvol}
  \frac{d\mathbb{P}^{u}}{d\mathbb{W}}=\exp\left\{\int_{0}^{t_n}\nu(u_s;\theta_x)du_s -\frac{1}{2}\int_{0}^{t_n}\nu(u_s;\theta_x)^2ds\right\},
\end{equation}
and the state vector can be written as
\begin{equation}
\label{eq:Vxtilde}
V_{0:n} =h_u(u_{0:n},x_0,\theta_x,\theta_v).
\end{equation}
The model can be defined from (\ref{eq:likeunitvol}), (\ref{eq:Vxtilde}) and (\ref{eq:PartObsDiffusion}). It contains intractable quantities but can be accurately approximated given the time discretisation of the diffusion path. An alternative reparametrisation, defined in discrete time, was suggested in \cite{Chib2006}. It uses the transformation below
\begin{equation}
  \label{eq:chibpitt}
w_t=\frac{x_t-\big(x_{t-\delta}-\delta\mu_x(x_t,\theta_x)\big)}{\sigma_{x}(x_{t-\delta},\theta_x)}, \forall t.
\end{equation}
In our setting the driving Brownian motion of $x_{0:n}^{dis}$, denoted by $w_{0:n}^{dis}$ and provided by \eqref{eq:chibpitt}, can be used to provide a discrete skeleton of the state vector
\begin{equation}
\label{eq:Vxtilde}
V_{0:n}^{dis}=h_w(w_{0:n}^{dis},x_0,\theta_x,\theta_v).
\end{equation}
The model is now given by $w_{0:n}^{dis}$, that can be transformed to $x_t$ for which the Euler-Maruyama approximation can be used, and \ref{eq:PartObsDiffusion} which can be approximated using the discretised state vector $V_{0:n}^{dis}$.

Data augmentations schemes can be used for the models above. Gibbs versions of such schemes will alternate between updating $u_{0:n}^{dis}$ (or $w_{0:n}^{dis}$) and consequently $V_{0:n}^{dis}$ given $\theta$, and $\theta$ conditional on either $u_{0:n}^{dis}$ (or $w_{0:n}^{dis}$). The first step can be done either by the overlapping block strategies in \cite{Chib2006} and \cite{Kalogeropoulos2007} or with a particle filter in the context of a particle Gibbs algorithm. The second step of updating $\theta_x$ given $u_{0:n}^{dis}$ or $w_{0:n}^{dis}$ is usually implemented through a random walk Metropolis algorithm:
\begin{itemize}
\item Let $\theta_x^c$ and $V_{0:n}^{disc}$ be the current values of $\theta_x$ and $V_{0:n}^{dis}$ respectively. Propose $\theta_x^*$ from $q(\theta_x^*|\theta_x^c)$.
\item Compute $V_{0:n}^{dis*}=h_u(u_{0:n},x_0,\theta_x^*,\theta_v)$
\item Accept with probability
$$
1\wedge\frac{\pi(\theta_x^{*},V_{0:n}^{dis*}|y_{1:n},u_{0:n}^{dis*}\theta_v,\theta_y)q(\theta_x^{c}|\theta_x^{*})} {\pi(\theta_x^{c},V_{0:n}^{disc}|y_{1:n},u_{0:n}^{disc}\theta_v,\theta_y)q(\theta_x^{*}|\theta_x^{c})}
$$
\end{itemize}
For the \cite{Chib2006} formulation, $u_{0:n}^{dis}$ can simply be replaced with $w_{0:n}^{dis}$ in the algorithm above.

\begin{figure}[h]
\begin{centering}
\includegraphics[scale=0.5]{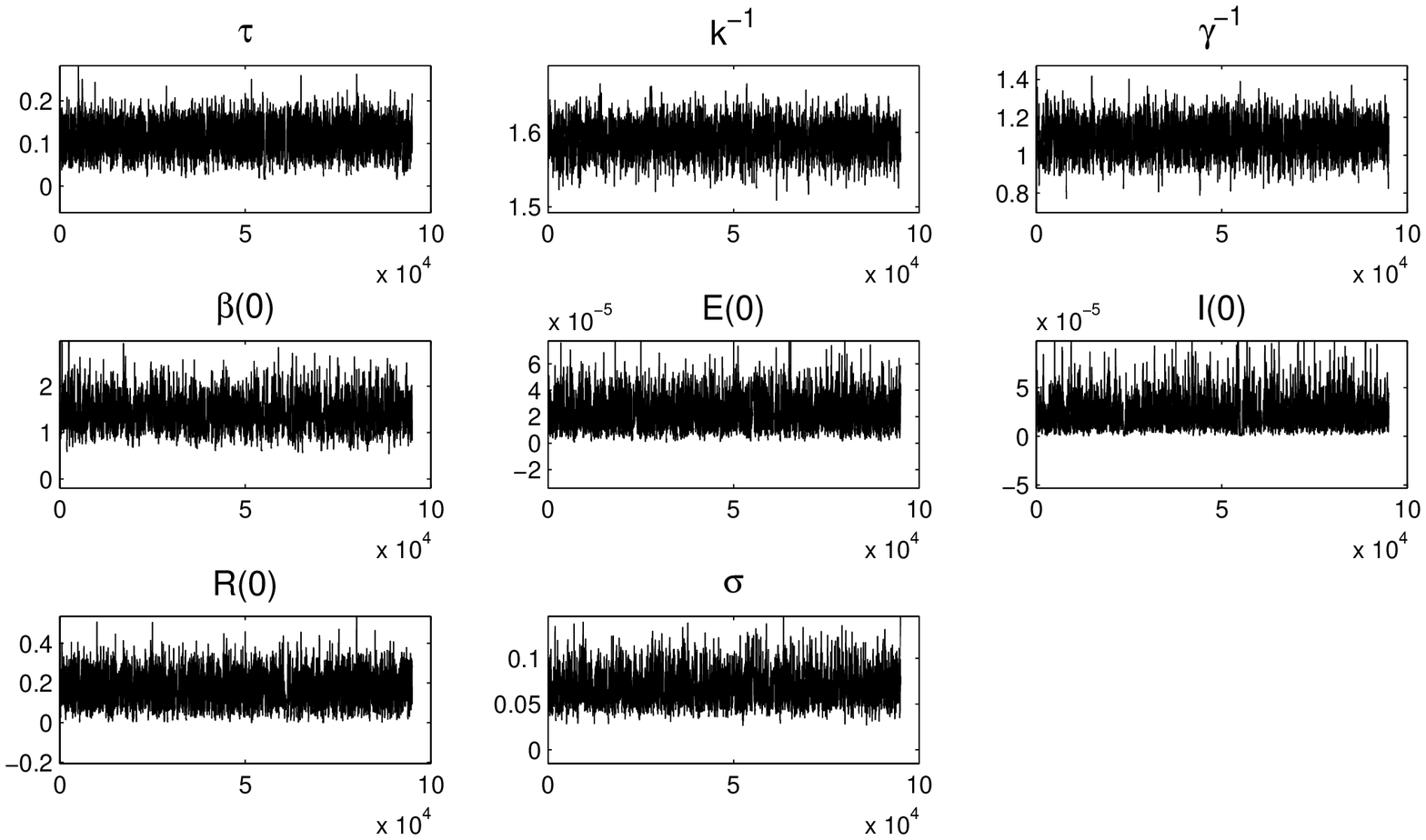}
\par\end{centering}
\caption{\label{fig:Trace} MCMC traceplots for each component of $\theta$}
\end{figure}

\begin{figure}[h]
\begin{centering}
\includegraphics[scale=0.5]{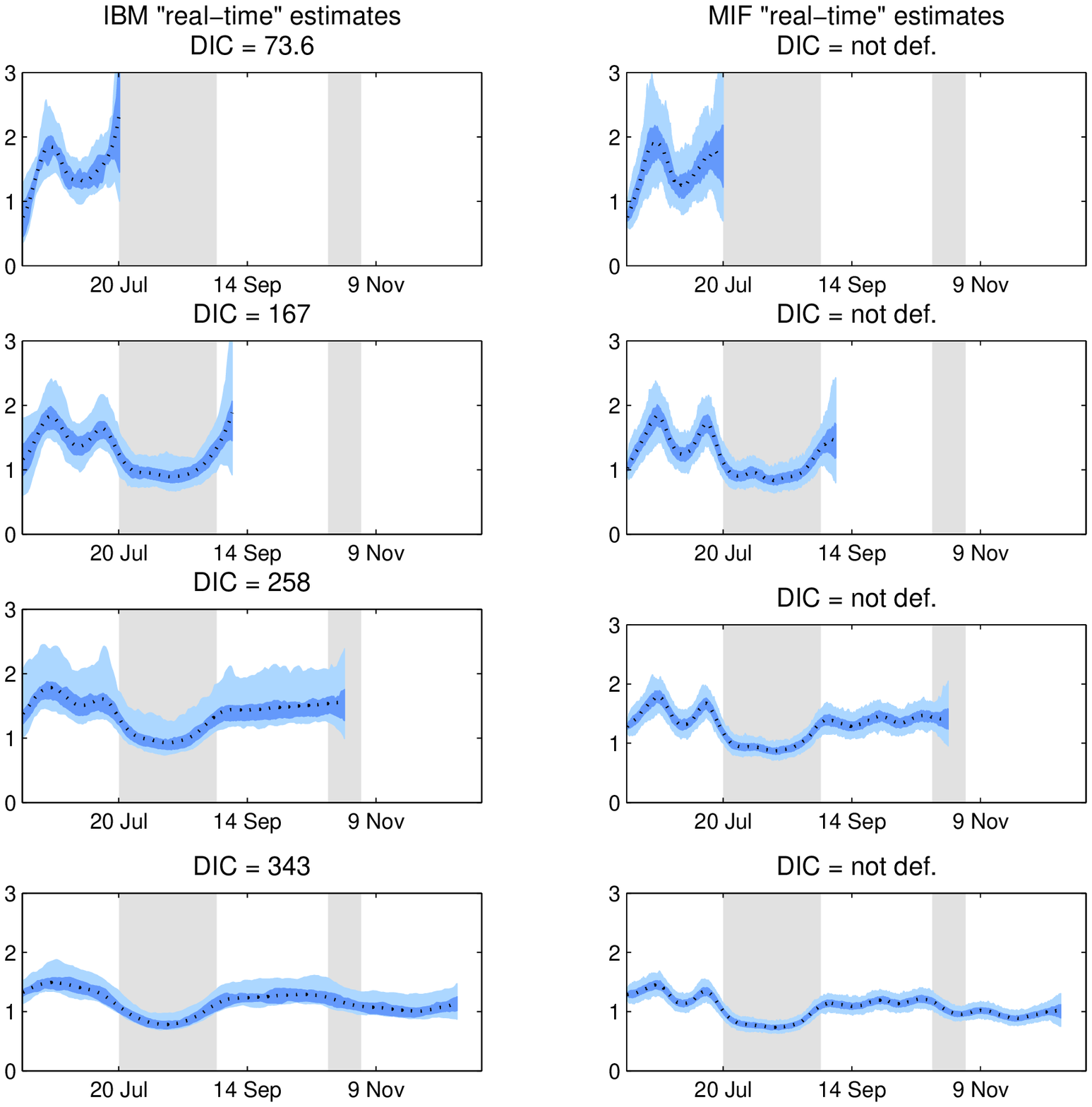}
\par\end{centering}
\caption{\label{fig:Comparing} Modeling choices and implications, aiming for robustness\protect \\
Black dotted lines indicate the mean of the pointwise posterior density. Dark and light blue show credible intervals, respectively at 50\% and 95\% levels.\protect \\
Left panels: estimates from an alternative modeling approach: exploring the full posterior density of an IBM diffusion model (left)\protect \\
Right panels: estimates from an alternative methodological approach: exploring the posterior density of a BM diffusion model conditionned on a likelihood maximizing parameter $\theta^*$ provided by the MIF algorithm (right)}
\end{figure}

\begin{figure}[h]
\begin{centering}
\includegraphics[scale=0.5]{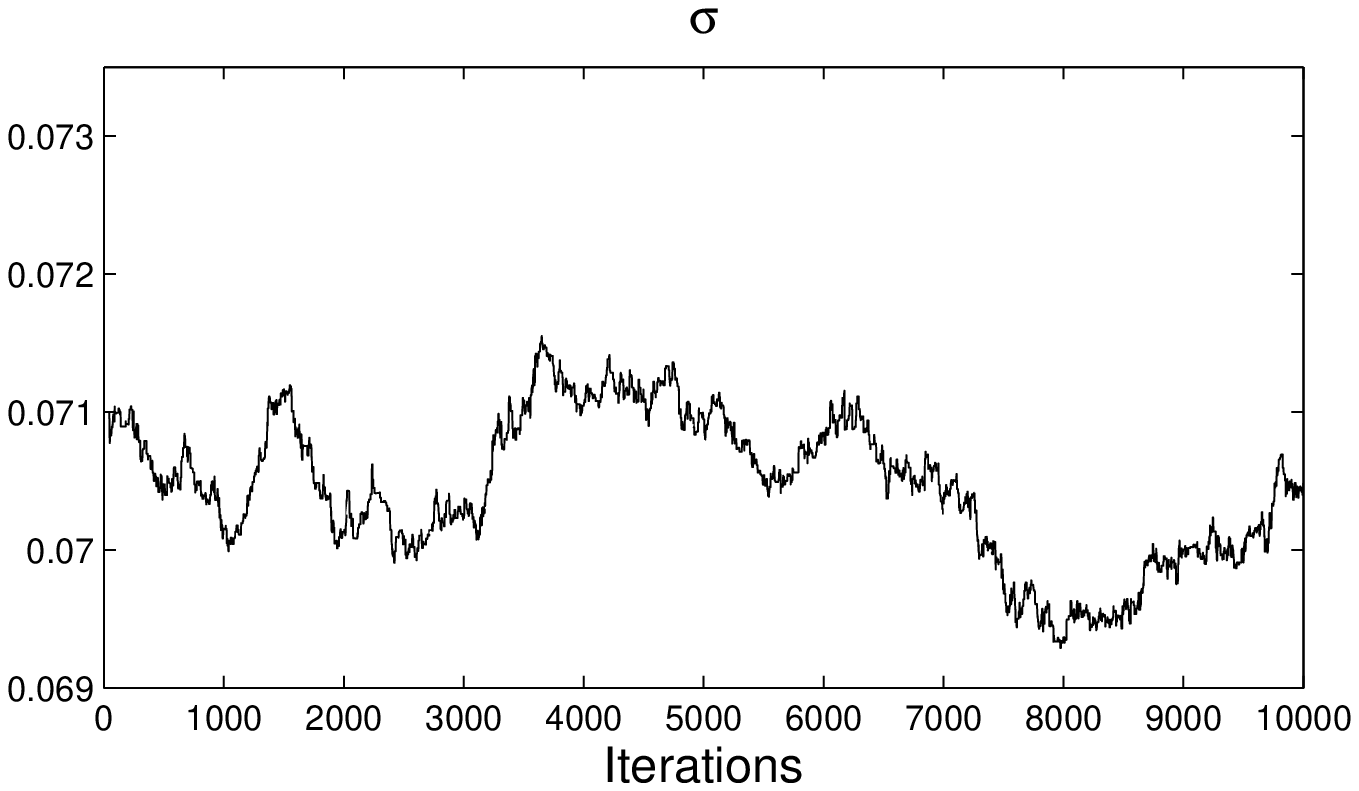}
\par\end{centering}
\caption{\label{fig:GibbsSigma} MCMC traceplot for $\sigma$ when using a Particle Gibbs scheme with reparametrization}
\end{figure}

Unfortunately both of the above algorithms may perform poorly. The problem is that every proposed value of $\theta_x$ implies a proposed trajectory of the ODE states vector $V_t$. As parts or functionals of this trajectory are observed with error, the proposed value of $\theta_x$ will not be accepted unless its associated $V_t$ trajectory is close to these observations. Consequently, only small steps can be made on the $\theta_x$ space and the algorithm mixes very slowly. The problem intensifies as the noise variance becomes smaller and as the time horizon of the epidemic increases. Implementations of such algorithms in the simulated and real data of this paper are in line with this argument. Figure \ref{fig:GibbsSigma} displays the posterior draws for $\sigma$ in the dataset of the simulation experiment 1 of Section 4 in the main paper. The posterior draws of $\sigma$ were obtained from a particle Gibbs algorithm combined with the algorithm above; note that in this model $u_{0:n}^{dis}$ and $w_{0:n}^{dis}$ are equal. In order to isolate the problem, the algorithm was run on $\sigma$ and $\beta_t$ only, and all the other parameters were held fixed at the values they where simulated from (a value of $\tau=0.1$ was used). The `true value' of $\sigma$ was $0.07$ and we used a $\delta=0.1$. As clearly shown in the traceplot the mixing of the chain is quite poor, thus casting doubts on the reliability of its output. The difference in mixing quality with the corresponding traceplot of Fig. \ref{fig:Trace} (bottom middle plot), corresponding to the PMCMC algorithm, is substantial.

To sum up, both formulations of Gibbs data augmentation schemes (with or without reparametrisation) are very likely to lead to inaccurate and inefficient MCMC algorithms. The use of particle MMH algorithms, termed as PMCMC in this paper, is therefore essential and the main paper focused on its implementation on diffusion driven epidemic models. PMCMC seems to provide a solution to the problem, but future research on Gibbs schemes with alternative reparametrisations addressing this problem, would be very helpful.

%
%

\bibliographystyle{biorefs}
\bibliography{Biblio}
\newpage

\begin{table}[h]
\begin{centering}
\begin{tabular}{ccc}
& \textbf{MSE} & \textbf{Bias}\tabularnewline
\hline
\textbf{EKF}&0.0269 & 0.0285\tabularnewline
\hline
\hline
\textbf{Particle filter}&0.0242 & 0.0064\tabularnewline
\hline
Improvement with regards to EKF& -10\% & -77\% \tabularnewline
\hline
\hline
\textbf{Particle smoother}& 0.0032& 0.0027\tabularnewline
\hline
Improvement with regards to P. filter& -87\% &-64\% \tabularnewline
\hline
\end{tabular}
\par\end{centering}
\caption{\label{table:MSEBias} Mean Squarred Error and Bias of $\beta_t$ estimates provided by the EKF, particle filter and particle smoother}
\end{table}

\begin{table}[h]
\begin{centering}
\begin{tabular}{cccccc}
& &  \textbf{Exp 1.a} & \textbf{Exp 1.b} & \textbf{Exp 1.b} & \textbf{Exp 2.b}  \tabularnewline
\hline
\hline
$\tau$ & Simulation value & \textbf{0.1} &  \textbf{0.05}  &  \textbf{0.1}  &  \textbf{0.05}  \tabularnewline
\hline
& Posterior mean & $0.103$ & $0.083$ & $0.078$ & $0.050$ \tabularnewline
\hline
& Posterior median & $0.103$ & $0.084$ & $0.077$ & $0.050$ \tabularnewline
\hline
& Posterior $95\%$ c.i. & $[0.051;0.152]$ & $[0.027;0.137]$ & $[0.063;0.96]$ & $[0.042;0.060]$ \tabularnewline
\hline
\hline
$\sigma$ & Simulation value &  \textbf{0.07} &  \textbf{0.07}  &  \textbf{n.d.}  &  \textbf{n.d.}  \tabularnewline
\hline
& Posterior mean & $0.066$ & $0.083$ & $0.016$ & $0.014$ \tabularnewline
\hline
& Posterior median & $0.064$ & $0.084$ & $0.015$ & $0.014$ \tabularnewline
\hline
& Posterior $95\%$ c.i. & $[0.048;0.090]$ & $[0.046;0.089]$ & $[0.010;0.027]$ & $[0.001;0.021]$ \tabularnewline
\hline
\end{tabular}
\par\end{centering}
\caption{\label{table:TauSigmaEsts} Mean, median and $95\%$ confidence intervals for $\tau$ and $\sigma$ estimates in four experiments}
\end{table}

\begin{table}[h]
\begin{centering}
\begin{tabular}{cccc}
\hline 
 & Posterior mean & Posterior median & Posterior 95\% c.i.\tabularnewline
\hline 
\hline 
$k^{-1}$ & $1.56$ & $1.55$ & $[1.53;1.60]$\tabularnewline
\hline 
$\gamma^{-1}$ & $1.00$ & $1.00$ & $[0.92;1.08]$\tabularnewline
\hline 
$\beta_{cc}(0)$ & $1.44$ & $1.36$ & $[0.89,2.30]$\tabularnewline
\hline 
$\beta_{aa}(0)$ & $1.40$ & $1.39$ & $[1.25;1.64]$\tabularnewline
\hline 
$\beta_{ca}$ & $0.30$ & $0.31$ & $[0.16;0.50]$\tabularnewline
\hline 
$\beta_{ac}$ & $0.32$ & $0.32$ & $[0.18;0.48]$\tabularnewline
\hline 
$E_{c}(0)$ & $2.1\times10^{-5}$ & $1.9\times10^{-5}$ & $[1.3\times10^{-5};3.8\times10^{-5}]$\tabularnewline
\hline 
$I_{c}(0)$ & $1.2\times10^{-5}$ & $1.4\times10^{-5}$ & $[0.6\times10^{-5};2.9\times10^{-5}]$\tabularnewline
\hline 
$R_{c}(0)$ & $0.13$ & $0.12$ & $[0.06;0.26]$\tabularnewline
\hline 
$E_{a}(0)$ & $2.1\times10^{-5}$ & $2.0\times10^{-5}$ & $[1.1\times10^{-5};3.5\times10^{-5}]$\tabularnewline
\hline 
$I_{a}(0)$ & $1.0\times10^{-5}$ & $1.0\times10^{-5}$ & $[0.3\times10^{-5};1.5\times10^{-5}]$\tabularnewline
\hline 
$R_{a}(0)$ & $0.16\times10^{-5}$ & $0.16$ & $[0.09;0.28]$\tabularnewline
\hline 
$\sigma_{c}$ & $0.11$ & $0.10$ & $[0.08;0.15]$\tabularnewline
\hline 
$\sigma_{a}$ & $0.08$ & $0.08$ & $[0.05;0.11]$\tabularnewline
\hline 
\end{tabular}
\par\end{centering}
\caption{\label{table:StructEsts} Mean, median and $95\%$ confidence intervals for the parameters of the structured model applied to the A/H1N1 pandemic data }
\end{table}

\begin{table}[h]
\begin{centering}
\begin{tabular}{ccccccccc}
\hline
 & $\tau$ & $k^{-1}$ & $\gamma^{-1}$ & $\beta_{0}$ & $E(0)$ & $I(0)$ & $R(0)$ & $\sigma$\tabularnewline
\hline
\hline
2.5\% quantile & 0.04 & 1.55 & 0.93 & 0.80 & $5.2\times10^{-6}$ & $1.6\times10^{-6}$ & 0.02 & 0.04\tabularnewline
\hline
25\% quantile & 0.09 & 1.57 & 1.03 & 1.16 & $1.6\times10^{-5}$ & $7.0\times10^{-6}$ & 0.12 & 0.05\tabularnewline
\hline
Median estimate & 0.11 & 1.59 & 1.08 & 1.35 & $2.3\times10^{-5}$ & $1.6\times10^{-5}$ & 0.17 & 0.06\tabularnewline
\hline
75\% quantile & 0.13 & 1.60 & 1.13 & 1.56 & $3.1\times10^{-5}$ & $2.8\times10^{-5}$ & 0.22 & 0.07\tabularnewline
\hline
97.5\% quantile & 0.17 & 1.63 & 1.23 & 2.13 & $5.2\times10^{-5}$ & 6.5$\times10^{-5}$ & 0.33 & 0.10\tabularnewline
\hline
Median when R(0) shifted +10\% & 0.11 & 1.59 & 1.09 & 1.41 & $2.0\times10^{-5}$ & $1.9\times10^{-5}$ & 0.19 & 0.06\tabularnewline
\hline
Median when R(0) shifted +20\% & 0.11 & 1.59 & 1.09 & 1.44 & 1.8$\times10^{-5}$ & $2.1\times10^{-5}$ & 0.24 & 0.06\tabularnewline
\hline
Median when R(0) shifted -10\% & 0.11 & 1.59 & 1.08 & 1.31 & $2.2\times10^{-5}$ & $2.2\times10^{-5}$ & 0.15 & 0.07\tabularnewline
\hline
Median when R(0) shifted -20\% & 0.12 & 1.59 & 1.09 & 1.27 & $1.9\times10^{-5}$ & $2.1\times10^{-5}$ & 0.13 & 0.07\tabularnewline
\hline
Median when $k^{-1}$ shifted +10\% & 0.11 & 1.59 & 1.09 & 1.33 & $1.8\times10^{-5}$ & $2.2\times10^{-5}$ & 0.15 & 0.06\tabularnewline
\hline
Median when $k^{-1}$ shifted +20\% & 0.12 & 1.60 & 1.08 & 1.34 & 1.8$\times10^{-5}$ & $2.3\times10^{-5}$ & 0.17 & 0.06\tabularnewline
\hline
Median when $k^{-1}$ shifted -10\% & 0.12 & 1.58 & 1.08 & 1.28 & $2.0\times10^{-5}$ & $2.0\times10^{-5}$ & 0.16 & 0.06\tabularnewline
\hline
Median when $k^{-1}$ shifted -20\% & 0.12 & 1.57 & 1.09 & 1.31 & $1.9\times10^{-5}$ & $2.0\times10^{-5}$ & 0.16 & 0.06\tabularnewline
\hline
Median when $\gamma^{-1}$ shifted +10\% & 0.11 & 1.59 & 1.12 & 1.23 & $2.2\times10^{-5}$ & $2.1\times10^{-5}$ & 0.15 & 0.07\tabularnewline
\hline
Median when $\gamma^{-1}$ shifted +20\% & 0.10 & 1.59 & 1.14 & 1.18 & $2.0\times10^{-5}$ & $2.4\times10^{-5}$ & 0.15 & 0.07\tabularnewline
\hline
Median when $\gamma^{-1}$ shifted -10\% & 0.11 & 1.59 & 1.06 & 1.37 & $1.9\times10^{-5}$ & $2.0\times10^{-5}$ & 0.16 & 0.07\tabularnewline
\hline
Median when $\gamma^{-1}$ shifted -20\% & 0.10 & 1.59 & 1.02 & 1.46 & $2.0\times10^{-5}$ & $1.8\times10^{-5}$ & 0.17 & 0.07\tabularnewline
\hline
\end{tabular}
\par\end{centering}
\caption{\label{table:Sensitivity} Original estimates compared to the ones resulting from respectively tilting the priors on $R(0)$, $\gamma^{-1}$ or $k^{-1}$ by +10, +20, -10 or -20\%}
\end{table}

